\documentclass[aps,showpacs,prd,twocolumn]{revtex4-1}%
\usepackage{graphicx}
\usepackage{amssymb}
\usepackage{amsmath}
\usepackage{color}
\usepackage{float}
\usepackage{tabularx}
\usepackage{ulem}
\usepackage{accents}
\usepackage{graphicx}
\usepackage{amsfonts}
\usepackage{wasysym}
\usepackage[toc,page]{appendix}
\usepackage{verbatim}

\newcommand{\be}{\begin{equation}}
\newcommand{\ee}{\end{equation}}
\newcommand{\bq}{\begin{eqnarray}}
\newcommand{\eq}{\end{eqnarray}}

\begin{document}

\title{Sorkin parameter for type-I SPDC biphotons and matter waves}

\author{F. C. V. de Brito  $^a$}\email[]{crislane.brito@ufabc.edu.br}
\author{C. H. S. Vieira  $^a$}\email[]{vieira.carlos@ufabc.edu.br}
\author{I. G. da Paz$^b$}\email[]{irismarpaz@ufpi.edu.br}
\author{J. B. Araujo $^c$}\email[]{jbaraujo@if.usp.br}
\author{M. Sampaio $^a$}\email[]{marcos.sampaio@ufabc.edu.br}

\affiliation{$^{a}$  CCNH, Universidade Federal do ABC,  09210-580 , Santo Andr\'e - SP, Brazil}

\affiliation{$^b$ Universidade Federal do Piau\'{\i}, Departamento de F\'{\i}sica, 64049-550, Teresina - PI, Brazil}

\affiliation{$^c$ Departamento de Física Matemática, Instituto de Física, Universidade de São Paulo, C.P. 66.318, São Paulo - SP, 05315-970, Brazil}

\begin{abstract}
\noindent
Type-I spontaneous parametric down-converted biphotons can be described approximately by a double-Gaussian wavefunction in configuration space. Using an effective propagator in the Fresnel approximation, the time evolution and transversal spreading of the two-particle biphoton wavefunction
allow us to evaluate the Sorkin parameter $\kappa$, which results from non-classical path contributions of kink-type and loops to  double and triple-slit interferometry. This simple unidimensional model for the evaluation of $\kappa$ predicts that kinked non-classical paths may lead to $\kappa \approx 10^{-5}$ for degenerate biphotons. We show that such a model reproduces well the Sorkin parameter for matter waves found in more involved approaches in the literature. Moreover we establish a hierarchy of  approximations based on the shape of the non-classical paths for matter waves and compare their size with leading relativistic corrections to the propagator.

\end{abstract}


\maketitle

\section{Introduction}

For over two hundred years Young's interference experiment has been crucial in probing nature's wave-particle duality. Interference phenomena allowed to set strong 
arguments in favor of the wave nature of light \cite{Young}, helped understanding the 
crystalline structure of materials \cite{crystal}, and showed that even large molecules, such as $C_{60}$, can behave like waves \cite{Fulerene,Viale} in benchtop experiments. Remarkably, interference has made it possible to verify the physical reality of the electromagnetic potentials \cite{AharonovBohm,Chambers}, rule out the existence of a luminiferous aether \cite{Morley}, and detect gravitational waves at the Laser Interferometer Gravitational-Wave Observatory (LIGO), in what is arguably the most precise scientific experiment in human history \cite{Ligo}.

\begin{figure}[h]
	\centering
	\includegraphics[width=5 cm]{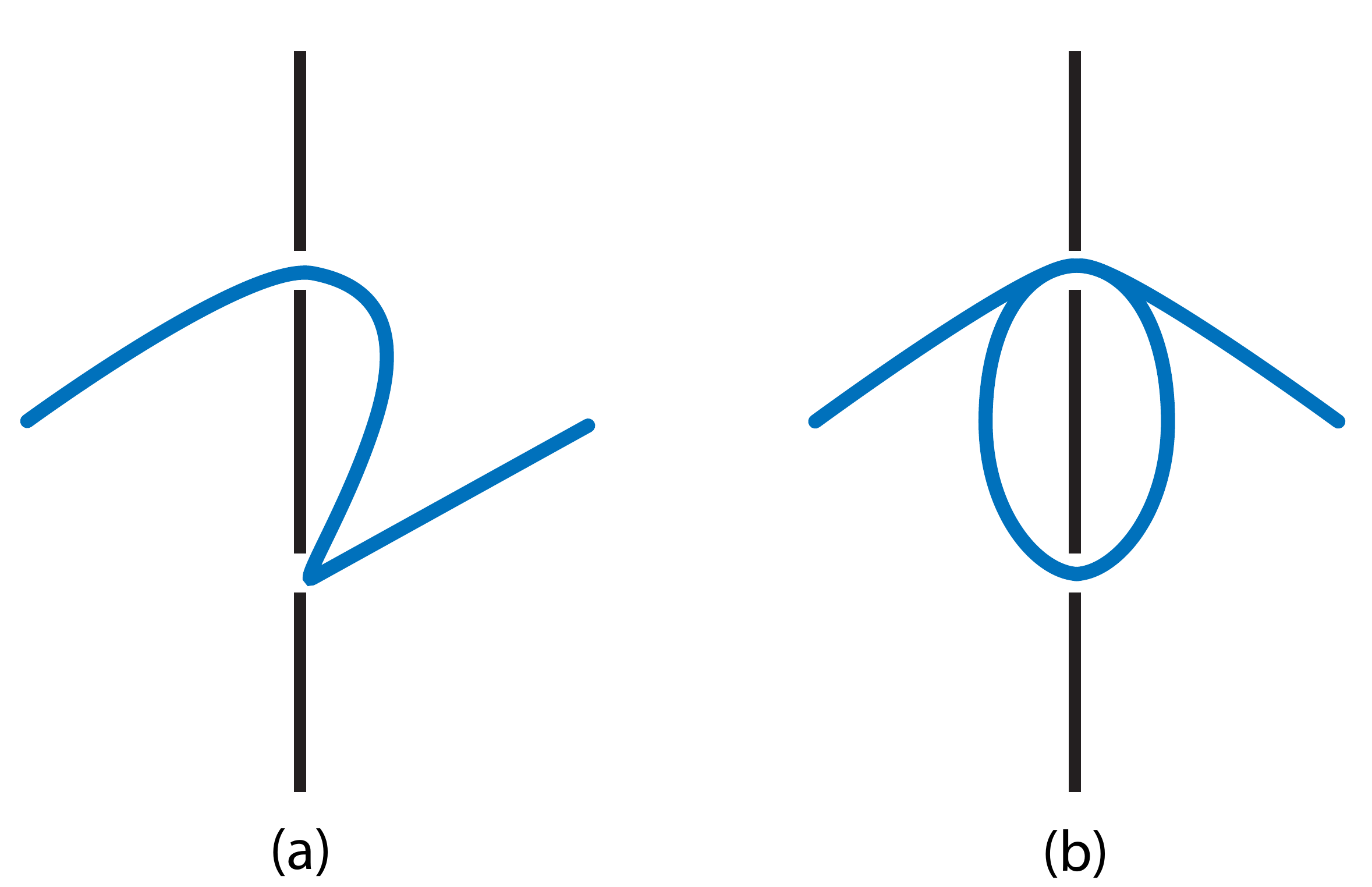}
	\caption{Lowest-order single-particle non-classical trajectories. (a) are referred to as kinks, while (b) as loops.}
	\label{KinkLoop}
\end{figure}

The most typical Young's experiment setup consists of a source, an opaque surface with two slits, and a screen at which the signal is detected. The Born rule states that if a quantum object is represented by a wave function $\psi(\vec{x}, t)$, than the probability density of detecting it at position $\vec{x}$ and time $t$ is given by the absolute square of the wave function \cite{Born}. In this away, when the standard superposition principle is applied in a double-slit experiment it has been common to consider that the wave function at the screen is a superposition of two amplitudes: one corresponding to the particle going through the upper slit, and the other, through the lower slit; these are usually called ``classical'' trajectories.
However, we run into trouble as the full problem (propagation through two simultaneously open slits) is not equivalent to the sum of those two possibilities (a single open slit at a time) -- these configurations do not possess the same boundary conditions. Of course, the problem is well posed. The probability amplitude for a particle to be at a space -time point $(\vec{x}_B,t_B)$ given that it started at $(\vec{x}_A,t_A)$ is given by the Feynman path integral
\begin{equation}
	\langle x_B|x_A \rangle = \int {\cal{D}} [x] \,\, e^{\frac{i}{\hbar} S[x]} ,
	\label{FPI}
\end{equation}
where $S[x]$ is the classical action, subjected to the constraints $x(t_A) = \vec{x}_A$ and $x(t_B) = \vec{x}_B$ \cite{Feynman}. For a potential representing a multi-slit barrier this is an overwhelmingly difficult problem even if treated numerically
\cite{Padmanabhan}. This poses an interesting and fundamental question: can we test the validity of the Born's rule and the superposition principle in multi-slit diffraction with light or matter waves? For this purpose one needs to consider leading non-classical (``exotic'' or sub-leading) trajectories that contribute to (\ref{FPI}) in a Young-type experiment. Yabuki \cite{Yabuki} was the first to exploit the contributions from such non-classical paths to the interference pattern in a double-slit experiment using both loops and kinks as shown in figure \ref{KinkLoop}.

For massive particles, two and three-dimensional models were implemented using a modified effective free particle propagator to account for the exotic paths. The effect of non-classical trajectories on the experiment's outcome is quantified by the Sorkin parameter $\kappa$, originally introduced in Ref. \cite{sorkin}. In a  multi-slit experiment, if $\psi_{A,B,C}$ represents the wavefunction at the detector for a particle emerging from slits $A,B,C$, the probability of detection at the screen is given by the Born rule:
\begin{eqnarray}
P_A &=& |\psi_A|^2, \nonumber \\
P_{AB} &=&  P_A + P_B + 2  \mathfrak{Re} (\psi_A^* \psi_B),\nonumber \\
P_{ABC} &=&  P_{AB} + P_{BC} + P_{AC} - P_A - P_B - P_C.
\end{eqnarray}
Notice that for three or more slits one always has a sum of terms denoting the interference of pairs of wavefunctions. A possible contribution from higher order terms is measured by 
\begin{eqnarray}
\varepsilon &=& P_{ABC} - P_{AB} - P_{BC} - P_{AC} + P_A + P_B + P_C, \nonumber \\
\kappa &\equiv& \varepsilon/I_{max},
\label{eqn:SORKINDEF}
\end{eqnarray}
in which the Sorkin parameter, here, has been  normalised with respect to the intensity at the central maximum $I_{max}$ in the interference pattern as defined in \cite{Sinha14}.

In \cite{Boyd16} the validity of Born's rule was verified through the experimental observation of exotic (looped) trajectories for the light by directly measuring their contribution to the formation of optical interference fringes in a triple-slit. Such exotic paths were enhanced with electromagnetic fields in the vicinity of the slits. The authors have verified that non-classical paths were related to the near-field component of the photon's wavefunction. Thus by controlling the strength and spatial distributions of the near fields around the slits, they claimed that the probability of looped trajectories were increased leading to $\kappa \approx |0.25|$ for x-polarised heralded photons of wavelength $810\,\, nm$ produced by degenerate down-conversion, in such a way that there was only one photon at a time in the experimental setup. The geometry involved  a triple-slit with height $h = 100\,\, \mu m$, slit width $w= 200\,\, nm$ and interslit separation $d= 4.6\,\, \mu m$. Conversely, $\kappa$ is almost zero when no enhancement was performed. By measuring each term in Eq. (\ref{eqn:SORKINDEF}), U. Sinha and collaborators \cite{Sinha2010} performed a three-slit experiment using different  photon sources such as an attenuated laser source down to $\approx 200\,\, fW$ and heralded single photons produced via spontaneous parametric down-conversion--SPDC of wavelength $810\,\, nm$. The typical sizes in their triple-slit apparatus were $h= 300\,\, \mu m$, $w = 30 \mu m$ and $d= 100 \mu m$. They determined a bound on the accuracy of Born's rule, namely that third order interference  was  less than $10^{-2}$ of the expected second order contributions given by the Born's rule. Moreover, semi-analytic and numerical methods were used in estimates for the Sorkin factor $\kappa$. For instance, in \cite{Sinha14}, an energy space propagator was used for both photons and electrons
\begin{equation}
K(\vec{r_1},\vec{r_2}) = \frac{k}{2 \pi i} \frac{e^{i k |\vec{r_1}-\vec{r_2}|}}{|\vec{r_1}-\vec{r_2}|},
\label{eqn:PROGSINHA}
\end{equation}
which satisfies the Helmholtz equation away from $\vec{r_1}=\vec{r_2}$ and the Fresnel-Huygens principle $K(\vec{r_1},\vec{r_3}) = \int d\vec{r_2} \, K (\vec{r_1},\vec{r_2}) K (\vec{r_2},\vec{r_3})$ for $\vec{r_2}$ integrated over a plane between $\vec{r_1}$ and $\vec{r_3}$ perpendicular to $\vec{r_1} - \vec{r_3}$. Such transitivity property is essential to write such a propagator in a path integral form \cite{Padmanabhan}
$$
K(\vec{r_1},\vec{r_2}) = \int {\cal{D}} [\vec{x}(s)] \exp [i k \int ds].
$$
where ${\cal{D}} [\vec{x}(s)]$ is the functional integration over the paths $\vec{x}(s)$ connecting $\vec{r_1}$ and $\vec{r_2}$. Thus, non-classical path contributions to $\kappa$ are numerically estimated in a triple-slit setup in the far-field (Fraunhofer) regime. In the thin-slit approximation, for incident photons of wavelength $\lambda= 810 \,\, nm$, $w = 30\,\, \mu m$ and $d = 100\,\, \mu m$, distance between source and slits and slits and detector equal to $18\,\, cm$, they found $\kappa \approx 10^{-6}$, whereas for electrons of $\lambda = 50\,\, pm$, $w = 62 \,\, nm$, $d = 272 \,\, nm $, source-slit separation $ 30.5\,\, cm$ and slit-screen distance $24\,\, cm$, $\kappa$ was estimated as $\approx 10^{-9}$. Within their model, they were able to verify that keeping other experimental
parameters fixed, $\kappa$ increases with an increase in $\lambda$ arriving at $\kappa \approx 10^{-3}$ for the microwave  regime and macroscopic distances such as $w = 1.2 \,\, m$ and $d = 4\,\, m$. Later on, an analytical description for the Sorkin parameter was derived and allowed for testing the r\^ole played by geometrical parameters on its determination \cite{Sinha15}. In that work, the authors obtain good agreement with the results of Ref. \cite{Sinha14} as well as with  sophisticated and enduring numerical finite-difference time-domain (FDTD) solutions of Maxwell's equations for realistic models of three-slit devices presented in \cite{FDTD}. In their analytical description for $\kappa$ using (\ref{eqn:SORKINDEF}) and the propagator (\ref{eqn:PROGSINHA}), successive approximations were possible assuming thin-slit and Fraunhofer limits (namely source-slit and slit-screen distances much greater than any other length scale). Moreover, in the Fresnel regime where such approximations are not valid, a $C^{++}$ code using Riemannian integration was used in \cite{Sinha15}. They have tested their code for the same parameters used for the photons in \cite{Sinha14} with a slit height  $h= 300 \,\, \mu m$. Starting with a slit-screen separation of $D = 20 \,\, cm$ which yields $|\kappa| \approx 6 \times 10^{-7}$, the value of $|\kappa|$ seems to increase as $D$ diminishes reaching a sudden peak at $D\approx 1.3\,cm$ which the authors attribute 
to a breakdown  of the paraxial approximation in the extreme near field regime. Another interesting breakthrough from the experimental viewpoint was reported in \cite{Sinha18}. Using a triple-slot experiment in the microwave domain, the authors  obtained a non-zero $\kappa$ using  a pyramidal horn antenna as a source of electromagnetic waves of $\lambda = \,\, 5 cm$ which reached on $10 \,\, cm$ wide slots and inter-slot distance $13 \,\, cm$. In addition, baffles were introduced inside the slits allowing for studying a hierarchy of subleading paths  contributing to $\kappa$.

Experiments testing the superposition principle to set bounds for the validity of Born's rule using massive particle multipath interferences were first performed in \cite{Cotter2017}. Cotter and collaborators  used a source of molecules with  $M= 515\,\, amu$ and Broglie wavelength $\lambda_{dB}= 2.5 \leftrightarrow 5.0 \,\, pm$. The diffraction mask was composed by single, double, and triple slits of width $w= 80 \,\, nm$ with periodicities $d= 100 \,\, nm$ and $d= 200 \,\, nm$ for the double  and $d= 100 \,\, nm$ for the triple-slit. In their experiment, a different definition of $\kappa$ was used, namely the normalization in Eq.  (\ref{eqn:SORKINDEF}) was taken with respect to the total number of molecules detected for a given Broglie wavelength leading to $|\kappa| \le 10^{-2}$. Likewise, metastable helium atoms were used in \cite{Barnea2018}. They have relied on a large number counting statistics ($1.7 \times 10^{6}$ counts in total) to obtain four diffraction patterns with a diffraction mask similar to \cite{Cotter2017} with $w= 84.5 \,\, nm$, $d= 136.5\,\, nm$ and $h= 1.6 \,\, mm$. The mask was placed at $\approx 60\,\, mm$ from the collimation device and $800 \,\, mm$ from the detection screen. Therefore, with that experiment, a new bound to  Born's rule using massive particle multipath diffraction was established at $|\kappa| \le 2.9 \times 10^{-5}$.

A simplified analytical model to evaluate the Sorkin parameter for matter waves was constructed in \cite{Paz2016,Geraldo2017,Vieira2019}. The authors consider a physical setup in which the quantum effects manifest chiefly in the transversal direction, say $\hat{x}$, alongside the slit widths and perpendicular to the momentum $\vec{p} = p_z \hat{z}$ of the particles emitted by the source. This turns out to be a good approximation in the limit where $\Delta p_z \ll p_z$,  allowing for treating the motion in the $z$-direction as classical since $p_z$ is sharply defined \cite{Viale}. The multi-slit interference pattern at the screen along the $x$-direction is obtained analytically through explicit integration using Gaussian shaped apertures. In order to assess the time spent by the particle during the inter-slit evolution of exotic paths the authors exploit the momentum uncertainty in the $x$-direction which is roughly $\epsilon \equiv m d /(\Delta p_x)$. In \cite{Paz2016} it was verified for electron waves that the Gouy phase difference $|\delta \mu_{G}|$ is due to  phases of non-classical path contributions. Thus $|\delta \mu_{G}|$ serves as a signal and measure of non-classical paths which led to $\kappa \approx 10^{-8}$ in a triple-slit construction  in accordance with \cite{Sinha14}. Using the same unidimensional model, a double-slit setup using two-level atoms and QED cavities positioned at the slit apertures was constructed in \cite{Geraldo2017}. The purpose was to account for the contribution of exotic trajectories only in the interference pattern via which-way information about the atoms. In this sense, in \cite{Quach2017} it was shown that non-classical paths yield different interference patterns using one and two which-way detectors in a double-slit experiment with light waves. This {\it{gedanken}} experiment was claimed to provide a new parameter (different from $\kappa$) to test  Born's rule, considering exotic paths as displayed in figure \ref{KinkLoop}(a) and the propagator \ref{eqn:PROGSINHA} in the Fraunhofer and stationary phase approximation. Finally in another contribution  \cite{Vieira2019} that employs the unidimensional model, a two slit experiment was modelled with cold neutrons using the following parameters: $m_n= 1.67 \times 10^{-27}\,\, kg$, $d= 125\,\, \mu m$, $w= 7 \,\, \mu m$, source-slit distance $z_T = 5.0 \,\, m$, slit-screen distance $z_\tau = 5.0 \,\,  m$, $\lambda_{dB} = 2 \,\, nm$, interslit propagation time $\epsilon = 19.6 \,\, ms$ leading to $\kappa \approx 10^{-5}$. Their analysis also allows for studying the behaviour of $\kappa$ with $z_\tau$ (Fresnel regime). Moreover the authors showed that the Sorkin parameter can be related to the visibility and axial phases (such as the Gouy phase) and thus they could be used as alternative quantifiers for exotic paths.

In this contribution we employ the one-dimensional model constructed in \cite{Paz2016,Geraldo2017,Vieira2019} to evaluate the Sorkin parameter. We address some questions related to the level of approximations involved in the analysis of kinked and looped non-classical paths for both matter and light particles using double and triple-slit constructions. The main advantage of this simplified model is that it is completely analytical and reproduces  the order of magnitude of Sorkin parameters computed with more sophisticated approaches. Furthermore we show that our approach can be extended to the effective biphoton wavefunction in the configuration space which describes twin photons produced by type I spontaneous parametric down-conversion (SPDC-I) \cite{OURPRA}. Because the signal of non-classical paths in the interference pattern is a relatively tiny effect, we assess the contribution of relativistic effects in the interference pattern to compare with exotic path contributions. Simple analytical methods are useful as they provide hints for experimentalists to detect such small effects.

This work is organised as follows. In section \ref{SectionII} we outline from first principles the approximations involved in the construction of biphoton wavefunctions as well as the entanglement measures and parameters encoded in a double Gaussian approximation. We also derive an effective propagator which describes the time evolution of the wavepackets and interference in the transversal direction. Section \ref{SectionIII} presents a consistency check of the framework in which it is verified that two entangled particles of wavelength $\lambda$ can behave  as a biphoton of wavelength $\lambda/2$. The Sorkin parameter $\kappa$ in a double and triple-slit setup is defined in section \ref{SectionIV} where  we also compute the leading order contributions to $\kappa$  for a biphoton produced via type-I SPDC. Moreover, within our framework we compute the leading order contributions of non-classical paths for matter waves (electron) and show that they are in agreement with more involved numerical and analytical methods in the literature. We finish that section by establishing a hierarchy of non-classical paths (kinks and loops) that contribute to $\kappa$  and compare their magnitude to relativistic corrections to the propagators. In Section \ref{SectionV} we draw our final remarks and conclusions and we leave the 
bulky formulae to the appendices.

\section{SPDC-I Biphoton  Wavefunction}
\label{SectionII}


A first quantised theory of a photon  is in principle not achievable because the photon is a massless relativistic quantum particle and thus is intrinsically described within (second quantised) quantum field theoretical formalism. Due to gauge symmetry the appropriate degree of freedom is the electromagnetic potential  $A_\mu$. This fact does not prevent us to: (a) describe the low intensity limit of a double-slit experiment with light within wave mechanics, (b) define an approximate position eigenstate for a photon nor (c)  investigate a quantum-mechanical description of optical beams \cite{Birula1}. One plausible approach due to J. R. Oppenheimer \cite{Oppenheimer1931} is based on an extension of the Weyl equation for massless neutrinos by replacing the Pauli vector with an angular momentum operator for spin-1 particles. A nice review can be found in  \cite{bookKeller}. The resulting six-component bispinors have positive and negative frequencies and can be interpreted as energy wavefunctions of photons and antiphotons, respectively. Moreover a Lorentz invariant measure for the scalar product can be defined as well as an approximate position state. However, the propagator and time dependent correlations within this approach is a tricky problem mainly due to the fact that photons are non-localizable. In \cite{Raymer} a modification in the Fourier transform in order to define the photon wavefunction was proposed. Whichever effective model one uses to describe a photon, it is important to take into account the process which generates it. In \cite{Field}  was presented a wave function description of a photon in Young double-slit experiment in which the photon source is a single excited atom (see also \cite{Keller}). Moreover, in \cite{Pablo} a second quantised version of the Bialynicki-Birula-Sipe photon wave function \cite{Birula2} formalism was
extended to include the interaction between photons and continuous (non-absorptive) media. As an application, 
the quantum state of the twin photons generated by SPDC was derived. That being said, an effective wavefunction treatment of photon states is possible and tools from Schroedinger wave mechanics may provide insights on various aspects of quantum light.

\begin{figure}[t]
	\centerline{\includegraphics[scale=0.4]{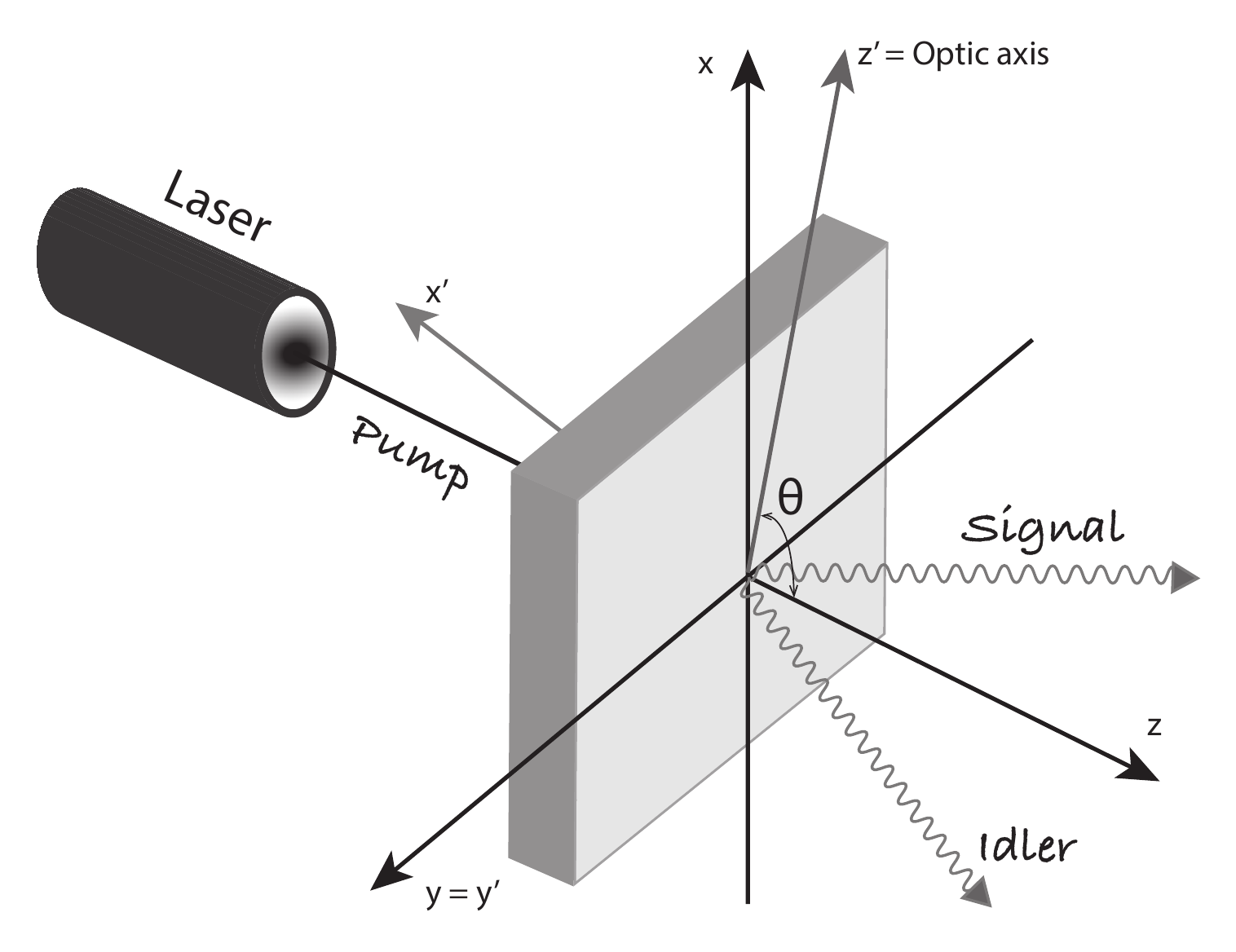}}
	\caption{Type I SPDC process. An uniaxial crystal of optical axis $z'$. Ordinary rays have polarization in a direction perpendicular to the plane $zz'$. Extraordinary rays have polarization on the plane $zz'$
		and experiences a refractive index $n_e (\theta)$ that depends on the angle $\theta$ between the optical axis and the longitudinal direction $z$ \cite{OURPRA}.}
	\label{fig:spdc}
\end{figure}

SPDC occurs when a nonlinear
and usually birefringent crystal is hit by an incoming
photon at (pump) frequency $\omega_p$ which in turn is
converted into two new outgoing photons of frequencies
$\omega_s$ (signal) and $\omega_i$ (idler) with  $\omega_p = \omega_i  + \omega_s $ and $\vec{k}_p = \vec{k}_i  + \vec{k}_s $. The polarization properties of the photon pair define the resulting spatial distribution and serve to characterise the SPDC phenomenon. A
type-I SPDC process happens when the polarization of
the outgoing photons is parallel to each other and orthogonal to the polarization of the incoming photon. The spatial distribution of the emerging photons forms a cone that is aligned with the pump beam propagation with the apex at the crystal (figure \ref{fig:spdc}).
The state of a down-converted photon pair may be constructed based on some reasonable simplifying assumptions \cite{OURPRA,Howell2016B,Walborn2010,Couteau2018}, such as that the crystal dimensions are large as compared to typical photon wavelengths, the crystal non-linear susceptibility tensor is a slowly-varying function of the frequencies,  the pump field is narrowband around $\omega_p$ and that its field amplitude does not vary significantly as it travels across the crystal. We can therefore write
\bq
|\Psi \rangle_{SPDC} &\approx& {\mathcal{C}}_0 |0_1, 0_2 \rangle + {\mathcal{C}}_1  \int_{\vec{\mathbf{k}}_1 \vec{\mathbf{k}}_2} \sqrt{\omega_{1} \omega_{2}} \Phi(\vec{\mathbf{k}}_1,\vec{\mathbf{k}}_2) \nonumber \\ &\times& \hat{a}^{\dagger}_{\vec{\mathbf{k}}_{1}}\,\hat{a}^{\dagger}_{\vec{\mathbf{k}}_{2}}|0_1, 0_2 \rangle ,
\eq 
where $1, 2$ are signal/idler photon indices, ${\cal{C}}_{0,1}$ are normalization constants, $\int_{\vec{\mathbf{k}}_i} \equiv \int d^3 \vec{\mathbf{k}}_i/(2 \pi)^3$ and
\bq
\Phi(\vec{\mathbf{k}}_{1},\vec{\mathbf{k}}_{2}) &=& \mathcal{N} \delta (\omega_{1} + \omega_{2} - \omega_p) \delta^2 ( \vec{\mathbf{q}}_1 + \vec{\mathbf{q}}_2 - \vec{\mathbf{q}}_p) \nonumber \\ &\times&
\text{sinc}\bigg(\frac{\Delta k_{z}L_{z}}{2}\bigg) \tilde{E}(\vec{\mathbf{q}}_{1}+\vec{\mathbf{q}}_{2}).
\label{wfmomentum}
\eq
In the equation above $\mathcal{N}$ is a normalization constant, $\Delta k_z \equiv k_{1z} +  k_{2z} -  k_{pz}$, $\vec{\mathbf{q}}_{i}$ are the momenta in the transversal direction, namely $\vec{\mathbf{k}} = (\vec{\mathbf{k}}^T, \vec{\mathbf{k}}^L) 
  \equiv (\vec{\mathbf{q}}, k_{z} \hat{z})$ 
and $L_z$ is the crystal thickness.

In addition as in the Fresnel (paraxial) approximation  $|\vec{\mathbf{q}}|^2 \ll |\vec{\mathbf{k}}|^2$, it is possible to express $k_z$ in terms of the transverse components $\vec{\mathbf{q}}$ \cite{OURPRA,Walborn2010} to yield
\bq
\hskip-0.4cm \Phi(\vec{\mathbf{q}}_{i},\vec{\mathbf{q}}_{s}) = \mathcal{N}_S
\; \text{sinc}\big(  b^2 |\vec{\mathbf{q}}_i -\vec{\mathbf{q}}_s|^2  \big) e^{-|\vec{\mathbf{q}}_i + \vec{\mathbf{q}}_s|^2/\sigma_{\perp}^2}
\label{wfmomentum2}.
\eq
In (\ref{wfmomentum2}), $b^2 \equiv \frac{L_z }{4 k_{ep}}$, $k_{ep}\equiv n_e(\theta) \omega_p/c$, $\mathcal{N}_s$ is the normalization and we assumed that the transverse pump momentum profile is Gaussian
$
\tilde{E} (\vec{\mathbf{q}}_i + \vec{\mathbf{q}}_s) = \widetilde{{\cal{N}}} e^{-|\vec{\mathbf{q}}_i + \vec{\mathbf{q}}_s|^2/\sigma_{\perp}^2}
$
which is nothing but a statement of the uncertainty in transverse momentum conservation. The $\vec{\mathbf{q}}_i - \vec{\mathbf{q}}_s$ argument in the sinc function expresses energy and (longitudinal) momentum conservation. Notice that $\Phi(\vec{\mathbf{q}}_{i},\vec{\mathbf{q}}_{s})$  is not separable into factors depending  on 
$\vec{\mathbf{q}}_{i}$ and $\vec{\mathbf{q}}_{s}$ and therefore  it is entangled (not factorable). 

In \cite{Eberly2004} it was shown that the degree of entanglement is governed by the product $\sigma_{\perp} b$. High entanglement is achieved when either $\sigma_{\perp} b \gg 1$ or $\sigma_{\perp} b \ll 1$, the minimum occurring for $\sigma_{\perp} b \approx 1$. Moreover  the sinc representation of the biphoton wavefunction is more entangled than its Gaussian approximation, which we shall discuss below, for the same values of $\sigma_{\perp} b$. Moreover  this biphoton wavefunction is approximately  separable \cite{Howell2016B} (subject to the paraxial approximation) into  a product of functions, one dependent on only x-coordinates, and the other dependent on only y-coordinates). That is because 
for small values of $x$ and $y$, $\text{sinc}(x+y)\sim \text{sinc}(x)\text{sinc}(y)$. In the paraxial approximation, the transverse momenta are much smaller than the pump momentum, and so the arguments of the sinc functions are very small ($\approx 10^{-3}$) \cite{Howell2016B}. Thus writing the $y$-component $q_{i\,y}, q_{s\,y}$ simply as $q_i, q_s$ yields
\be
\Phi_S (q_i, q_s) = \widetilde{\mathcal{N}}_S
\; \text{sinc}\big(  b^2 (q_i - q_s)^2  \big) e^{-(q_i+q_s)^2/\sigma_{\perp}^2}.
\label{wfmomentum4s} 
\ee
In order to study the spatial transversal correlations of the biphotons we need to Fourier transform the wavefunction into coordinate space. Following \cite{Howell2016B,Walborn2010,Eberly2004,Monken1999,Fedorov2009,Qureshi2018,Bramon2005} the sinc function may be approximated by a Gaussian
\be
\Phi_G (q_i, q_s) =  \widetilde{\mathcal{N}}_G e^{-b^2 (q_i - q_s)^2} e^{-(q_i+q_s)^2/\sigma_{\perp}^2},
\label{wfmomentum4g}
\ee
whose Fourier transform is
\be
\Psi_G (y_-,y_+) = \frac{1}{\sqrt{2 \pi \sigma_- \sigma_+}} \; e^{-\frac{y_-^2}{4 \sigma_-^2}} \; e^{- \frac{y_+^2}{4 \sigma_+^2}},
\ee
where $\sigma_- \equiv b/\sqrt{2}$, $\sigma_+ \equiv \sqrt{2} \sigma_{\perp}$ and  $y_{\pm} \equiv \frac{(y_i \pm y_s)}{\sqrt{2}}.$ Also we have normalised $\Psi_G$ so that $\int \int dy_- \; dy_+ |\Psi_G|^2 = 1$. We shall adopt the double-Gaussian approximation for simplicity as it makes both transverse position and momentum statistics easy to calculate besides fitting well experimental data \cite{ExpDataGauss}. Moreover, as we shall see, the double-Gaussian wavefunction is easy to propagate in time within the paraxial regime (the same regime used in the approximations of our biphoton state).

Consistently with the approximations that led to the down-converted biphoton wavefunction, under the conditions of validity of the Fresnel approximation, the diffraction and interference of a wave travelling in the $z$-direction can be described in terms of its spreading in time of the  wavepacket transversal $(x,y)$-section \cite{Dillon2011}. For Broglie waves of massive particles, the wavepacket spreading is due to the dispersion relation $\omega_k = \hbar k^2/(2 m)$
and the free evolution is  given by the Fourier transform 
\be 
\psi (\vec{r},t)= \int d^3 k \; e^{(i \vec{k}.\vec{r} - i \omega_k t)} \; \tilde{\psi}(\vec{k},0), 
\ee
where $\tilde{\psi}(\vec{k},0)$ is the Fourier transform of the initial condition. As for  a biphoton wave travelling in the $z$-direction, due to the fact that the sinc function factorises in the transversal $(x,y)$ directions for typical experimental parameters, we may disregard the $x$-direction. In the case of a multi-slit diffraction, we could consider such waves impinging on a screen with slits along $x$-axis and study the spreading along the $y$-axis. Thus assuming symmetry along the $x$-axis, we may disregard the $x$-coordinate and write \cite{Dillon2011}
\be
\Psi (y,z,t) = \psi (y,z) e^{-i \omega_0 t},
\ee
in which $\psi$ satisfies the Helmholtz equation $\triangle \psi = -k_0^2 \psi$ and $\omega_0 = c k_0$. Taking the one-dimensional Fourier transform:
\be
\psi (y,z) = \frac{1}{\sqrt{2 \pi}} \int \tilde{\psi}(k_y,z) e^{i k_y y} \; dk_y,
\ee
and using that $\psi (y,z)$ satisfies the Helmholtz equation we get, for progressive waves in the $z$-direction,
\be
\tilde{\psi}(k_y,z) = \tilde{\psi}(k_y,0) e^{i \sqrt{k_0^2-k_y^2} z}
\ee
which, in the Fresnel  approximation $\sqrt{k_0^2-k_y^2} \approx k_0 - k_y^2/(2 k_0)$, gives \cite{Dillon2011}:
\be
\psi (y,z) = e^{i k_0 z} \frac{k_0}{\sqrt{2 \pi i z}} \int e^{i \frac{k_0}{2 z} (y-y')^2} \psi (y',0).
\ee
By identifying $\psi (y, t=0) \equiv \psi (y, z=0)$ we have  $|\psi (y,t)|^2 \equiv |\psi (y, z)|^2$ provided 
$z = c t$. Therefore we arrive at the nonrelativistic-like propagator:
\be
G (y , t ; y', t') = \sqrt{\frac{\tilde{m}}{2 \pi i \hbar (t - t')}} e^{i \frac{\tilde{m} (y - y')^2}{2 \hbar (t - t')}},
\label{propagator}
\ee
where $
\tilde{m} \equiv \frac{k_0 \hbar }{c} $ and we have dropped out a global phase factor $e^{-i \frac{\tilde{m} c^2}{\hbar} (t - t')}$ which is immaterial.
The propagator (\ref{propagator}) was used in \cite{Qureshi2018} in a double-slit experiment to demonstrate that a degenerate biphoton of wavelength $\lambda$ produced via SPDC can behave as a single quanton of wavelength $\frac{\lambda}{2}$ as seen in \cite{Monken1999}. It was also employed in \cite{OURPRA} to study a continuous-variable Bell violation for type I-SPDC biphotons. Finally, we write the free propagation of a biphoton SPDC wavefunction  as
\bq
\Psi (y_i,y_s,t) &=& \iint d y_i' \, dy_s' \;  G (y_i , t ; y_i', 0) G (y_s , t ; y_s', 0) \nonumber \\ &\times& \Psi (y_i',y_s',0).
\label{eqn:propbiphoton}
\eq 
To make contact with the notation in the literature, let us redefine the biphoton coordinates such that 
$$
y_i \equiv x_1, \,\, y_s \equiv x_2, \,\, \sigma_- \equiv \sigma/\sqrt{2} \,\, \text{and} \,\, \sigma_+ \equiv \Omega/\sqrt{2} \
$$
and therefore
\begin{equation}
\psi_0 (x_1,x_2)=\frac{1}{\sqrt{\pi \sigma \Omega}} e^{-\frac{(x_1 - x_2)^2}{4 \sigma^2}} e^{-\frac{(x_1 + x_2)^2}{4 \Omega^2}},
\label{estadoinicial}
\end{equation}
as well as new relative coordinates 
$r=(x_1+x_2)/2$ and $q=(x_1-x_2)/2$, so that 
\begin{equation}
 \psi_0 (r,q)=\frac{1}{\sqrt{\pi \sigma \Omega}} e^{-\frac{q^2}{\sigma^2}} e^{-\frac{r^2}{\Omega^2}}.
\label{estadoinicial}
\end{equation}
Accordingly, after a time $t$, the evolved state in the $\{r,q\}$ variables becomes, using (\ref{eqn:propbiphoton})
\begin{equation}\label{psifp}
 \begin{split}
\psi (r,q,t)=C \exp\bigg[\frac{-q^2}{\sigma^2+\frac{i \lambda c t}{2 \pi}}\bigg] \exp\bigg[\frac{-r^2}{\Omega^2+\frac{i \lambda c t}{2 \pi}}\bigg],
\end{split}
\end{equation}
in which
\begin{equation}
C=\frac{1}{\sqrt{\pi\left[\sigma+i(\frac{ \lambda c t}{2 \pi})\frac{1}{\sigma}\right]\left[\Omega+i(\frac{ \lambda c t}{2 \pi})\frac{1}{\Omega}\right]}}.
\end{equation}

\begin{figure}[h]
\centering
\includegraphics[width=8.5 cm]{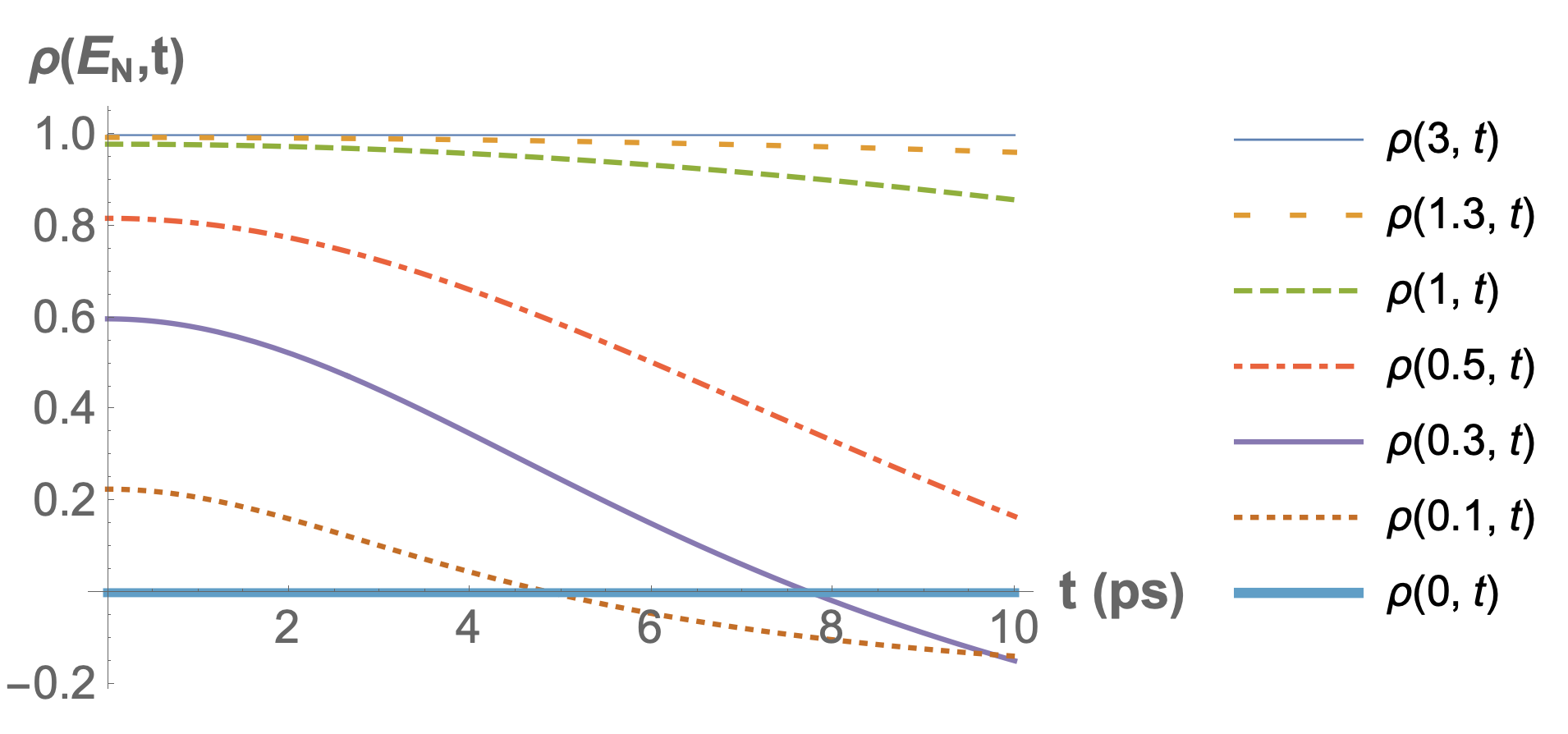}
\caption{Behaviour of the degree of spatial correlation as a function of $t$ (in picoseconds) for different values of logarithmic negativity $E_{\mathcal{N}}$. One observes that for $E_{\mathcal{N}}\sim3$, the initial value of $\rho(t)$ saturates at $1$ and barely decreases for $t=4\,$ps, which is the typical time it takes for the photons to reach the slits. For lower values of $E_{\mathcal{N}}$, the decrease can be appreciable. We have used $\sigma=11.4\times10^{-3}\,$mm, $c=0.3\,$mm/ps and $\lambda=7.02\times10^{-4}\,$mm.} \label{Rho}
\end{figure}

In order to characterise the entanglement of the transverse canonical coordinates $x_i$ and $p_i$ for the biphotons we follow \cite{ford, dorlas, ostermeyer}. The degree of entanglement can be quantified in the double Gaussian approximation in terms of the negativity of the partially transposed density matrix. The Duan criterion \cite{Duan} is a sufficient criterion for non-separability for a pair of EPR-type wavefunctions for continuous variables. For the wavefunction (\ref{psifp}) it yields that the system is separable if $\sigma =\Omega$. Also, the Peres-Horodecki criterion \cite{Simon} states that a Gaussian state is separable if and only if the minimum value of the  symplectic spectrum of the partial transpose of the covariance matrix is greater than $1/2$ which leads to a measure of the entanglement $E$ of the Gaussian state (\ref{psifp})
\be
E = \log_{10}	 \Big( \frac{\Omega}{\sigma} \Big) \equiv E_{\cal{N}},
\ee
that coincides with the expression for the logarithmic negativity $E_{\cal{N}}$ \cite{ford} that establishes that the greater $\Omega/\sigma$, the larger the negativity and hence the larger the entanglement. Another useful quantity is the degree of spatial correlation (Pearson $r$-value),
\be
\rho_x = \frac{\langle x_1 x_2 \rangle - \langle x_1\rangle \langle x_2 \rangle}{\sigma_{x_1}\sigma_{x_2}},
\ee
which ranges from $-1$ to $+1$,  where $\sigma_{x_{1,2}}$ is the standard deviation of $x_{1,2}$.  $\rho_x$ is zero if the two photons are uncorrelated, and $\rho_x \rightarrow +1$ if they are spatially closely correlated (bunched) and $\rho_x \rightarrow -1$ if they are spatially closely anti-correlated (anti-bunched). For the biphoton state described in (\ref{psifp}), we get
\begin{equation}
\rho_x(t)=\frac{\left(\Omega^{2}-\sigma^{2}\right)}{\left(\Omega^{2}+\sigma^{2}\right)}\frac{\left[1-\left(\frac{\lambda ct}{2\pi\sigma\Omega}\right)^{2}\right]}{\left[1+\left(\frac{\lambda ct}{2\pi\sigma\Omega}\right)^{2}\right]}.
\label{pearson}
\end{equation}
Accordingly, it is possible to write the degree of spatial correlation $\rho(t)$ as a function of the logarithmic negativity $E_{\mathcal{N}}$ and time $t$ -- a few plots of $\rho(E_{\mathcal{N}},t)$ are shown in figure \ref{Rho}.
Transverse spatial correlations of biphotons produced via SPDC in the double Gaussian approximation were studied in \cite{Howell2016B}. At the sinc-level, spatial correlations of biphotons were seen to encompass Bell non-locality in \cite{OURPRA}. In \cite{Qureshi2018}, a double Gaussian approximation was used to show a experimental verification \cite{Monken1999,Jacobson} that two entangled photons of wavelength $\lambda$ can behave like a single ``quanton'' of wavelength $\lambda/2$. Position-momentum Bell non-locality via a Clauser-Horne-Shimony-Holt inequality violation using entangled biphotons has also been  verified experimentally  in \cite{Schneeloch2016}. The entanglement of degenerate type-I SPDC biphotons was studied using a spectral wavefunction beyond double Gaussian approximation in \cite{Mikhailova2008},

Biphoton double-slit interference has been studied both theoretically and experimentally in a series of articles: in \cite{Hong1998} it was reported a non-local interference between SPDC biphotons measured in coincidence after passing through double-slits. Similarly, the interference pattern of two indistinguishable photons sent to well-defined slits at an identical time was analysed in \cite{brida}. Their data were in accordance with predictions based on standard quantum mechanics and in contrast with the deterministic Broglie-Bohm model. The role of mode functions and ``which-slit'' information in interference patterns of biphotons was experimentally assessed in \cite{Menzel}. Moreover, Young's double-slit interference with two-color biphotons was performed in \cite{2Color}, shedding further light on the interplay between interference and which-path information as a result of the nonlocal nature of two-photon entanglement.

\section{Biphoton interference in a double-slit}\label{SectionIII}

Considering only classical trajectories the biphoton state at the screen, after passing through a double-slit, can be written under the assumptions that led to (\ref{propagator}) as
\bq
\label{integral}
&&\psi(x_1, x_2, T + \tau)=\int_{X} G(x_1,  T+\tau;x_1'', T)\times 
\nonumber \\ && \times G(x_2,T+\tau; x_2'', T) 
F_{u,d}(x_1'')F_{u,d}(x_2'') \times \nonumber \\ &&
\times G(x_1'',T;x_1',0) G(x_2'',T;x_2', 0)   \psi_0(x_1',x_2'),
\eq
where the integrations from $-\infty$ to $+\infty$ are  taken over $\{X\}=\{x_1',x_2',x_1'',x_2''\}$. The functions $F$ represent the Gaussian-shaped windows \cite{Feynman}, which crop the wave function at the slits, and $ T (\tau)$ is the time interval between source and slits (slits and screen). The window functions $F_{u,d}(x_1'')F_{u,d}(x_2'')$ read
\be
\label{frq1}
F_u(x_i) \equiv e^{-\frac{( x_i - d/2 )^2}{2\beta^2}} \,\,\,\,\, \text{and}\,\,\,\,\,
F_d(x_i) \equiv e^{-\frac{( x_i + d/2 )^2}{2\beta^2}},
\end{equation}
where $i=1, 2$, $u(d)$ stands for upper(lower)-slit, $d$ is the interslit center-to-center distance and $\beta$ the slit width.
The integrals in Eq. (\ref{integral}) can be analytically computed to yield four  amplitudes
\begin{equation}
\psi_{i}=A_i\exp{\left[C_{i}(r,q) + i \alpha_i(r,q)\right]},
\end{equation}
with $C_{i}(r,q)$ and $\alpha_i(r,q)$ $\in \mathbb{R}$ and $i=\{uu,dd,ud,du\}$ denote each of the biphoton possible paths through the upper or lower slit. The coefficients $C_{i}$ and $\alpha_{i}$ have the general forms
\begin{align}
C_i(r,q)&=c_1 r^2 +c_2 q^2+c_3 r+c_4 q+c_5\nonumber\\
\alpha_i(r,q)&=a_1 r^2 +a_2 q^2+a_3 r+a_4 q+a_5,\label{alphaC}
\end{align}
whose coefficients $c_i$ and $a_i$ are listed in appendix A. The intensity at the screen is given by Born's rule
\begin{equation}\label{it}
I(x_1,x_2)=|\psi_{uu}+\psi_{ud}+\psi_{du}+\psi_{dd}|^2 = I(x_2,x_1)
\end{equation}
as we have considered biphotons such that $\lambda_1 = \lambda_2$ (degenerate case).

\begin{figure}[h]
\centering
\includegraphics[scale=.3]{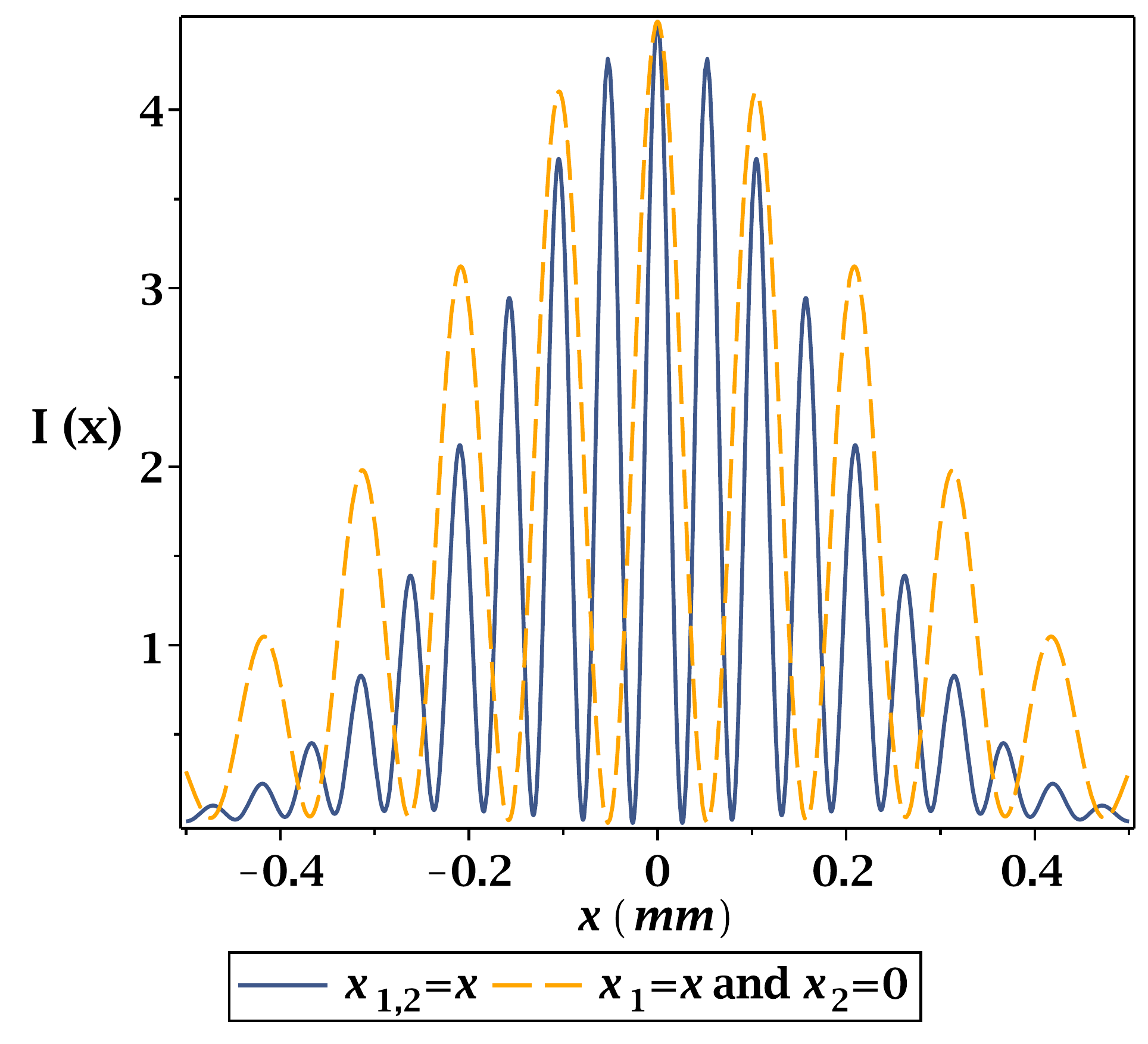}
\caption{Two photons of wavelength $\lambda$ behaving as a single photon of wavelength $\lambda/2$. We used the set of parameters: $\lambda=7.02\times10^{-4}\,$mm, $E_\mathcal{N}=2$, $\sigma=11.4\times10^{-3}\,$mm, $\Omega= 10^{E_\mathcal{N}}\sigma$, $\beta=5\times10^{-3}\,$mm, $d=0.1\,$mm, $c=0.3\,$mm/ps, $t=4\,$ps and $\tau=50\,$ps. These parameter values do not differ considerably from the ones used in Ref. \cite{brida}.}
\label{Quanton}
\end{figure}

At this point, it is worthwhile to test our framework by verifying that the diffraction of a wave-packet of two entangled photons, each of which of wavelength $\lambda$, can display an interference pattern of a single ``quanton'' of wavelength $\lambda/2$ \cite{Monken1999,Qureshi2018}. The values of $\Omega$ and $\sigma$ are 
determined by the experiment. Typically $\sigma = \sqrt{L_z \lambda_p/(6 \pi)} \approx 0.01 \, mm$ \cite{Howell2016B,Schneeloch2016}. Choosing 
 $E_\mathcal{N}=2$ (and thus high spatial correlation, see figure \ref{Rho}). For detection in coincidence  $x_1 = x_2 =x$ (or equivalently $r=x$ and $q=0$), we get the solid blue line in figure \ref{Quanton}. To verify whether a photon of the pair behave as a single particle, we may place one detector at the center of the slits ($x_2=0$) and let the other sweep the screen to obtain the intensity depicted by the dotted line in figure \ref{Quanton}. It is clear that the intensity for both photons detected at the same point oscillate with half the wavelength as compared with the single photon interference. This is in agreement with  \cite{Jacobson} in the sense that $N$ particles of wavelength $\lambda$ can behave like a single particle, or quanton, of wavelength $\lambda/N$ in an interference experiment. For entangled biphotons this has been verified in \cite{Walborn2010}. Moreover, in \cite{Qureshi2018} it was argued that this effect can be verified in a non-local fashion as well.
 In our case the high degree of spatial correlation at the slits ($T\approx 4 \, ps$) turns  the amplitudes $\psi_{ud,du}$ highly suppressed, namely the wavefunctions $\psi_{ud,du}$ can be removed from the total intensity Eq. (\ref{it}) with negligible effect (the fractional difference is of order $\sim 10^{-15}$), which indicates the photons are likely to go through the same slit. A similar analysis  was performed in \cite{Qureshi2018} considering sharp slits and ignoring the  middle terms in (\ref{it}) for high spatial correlation. 

\section{Sorkin Parameter}
\label{SectionIV}

Let us evaluate the Sorkin parameter due to the biphoton's non-classical trajectories in double and triple-slit setups.  The Sorkin parameter can be defined as
\begin{equation}
\kappa=\frac{I_\text{total}-I_\text{c}}{I_{\text{total}}(0)}. \label{sorkinDef2}
\end{equation} 

For a single particle triple-slit setup, according to the Born's rule, the  probability of detection at the screen is given by%
\begin{equation}
I_{\text{total}}=I_{ABC}=\vert \psi_A+\psi_B+\psi_C+\psi_{\text{nc}}\vert^2,
\label{eqn:BR},
\end{equation}
where $I_\text{c}=\vert \psi_A+\psi_B+\psi_C\vert^2$, $\psi_i$ is for the wavefunction at the screen when a particle passes through slit $i$, and $\psi_{\text{nc}}$ corresponds to any exotic, non-classical trajectories.

For a biphoton double-slit setup, the total intensity is
\begin{equation}
I_\text{total}=\vert\psi_{uu}+\psi_{ud}+\psi_{du}+\psi_{dd}+\psi_\text{nc}\vert^2
\label{IntensityNC},
\end{equation}
while the classical contribution reads
\begin{equation}
I_\text{c}=\vert\psi_{uu}+\psi_{ud}+\psi_{du}+\psi_{dd}\vert^2.
\end{equation}
Next we describe the non-classical trajectories and rank them according to their contribution to the Sorkin parameter.

\subsection{Sorkin parameter for the biphoton: double-slit}

\begin{figure}[h]
\centering
\includegraphics[width=5.0 cm]{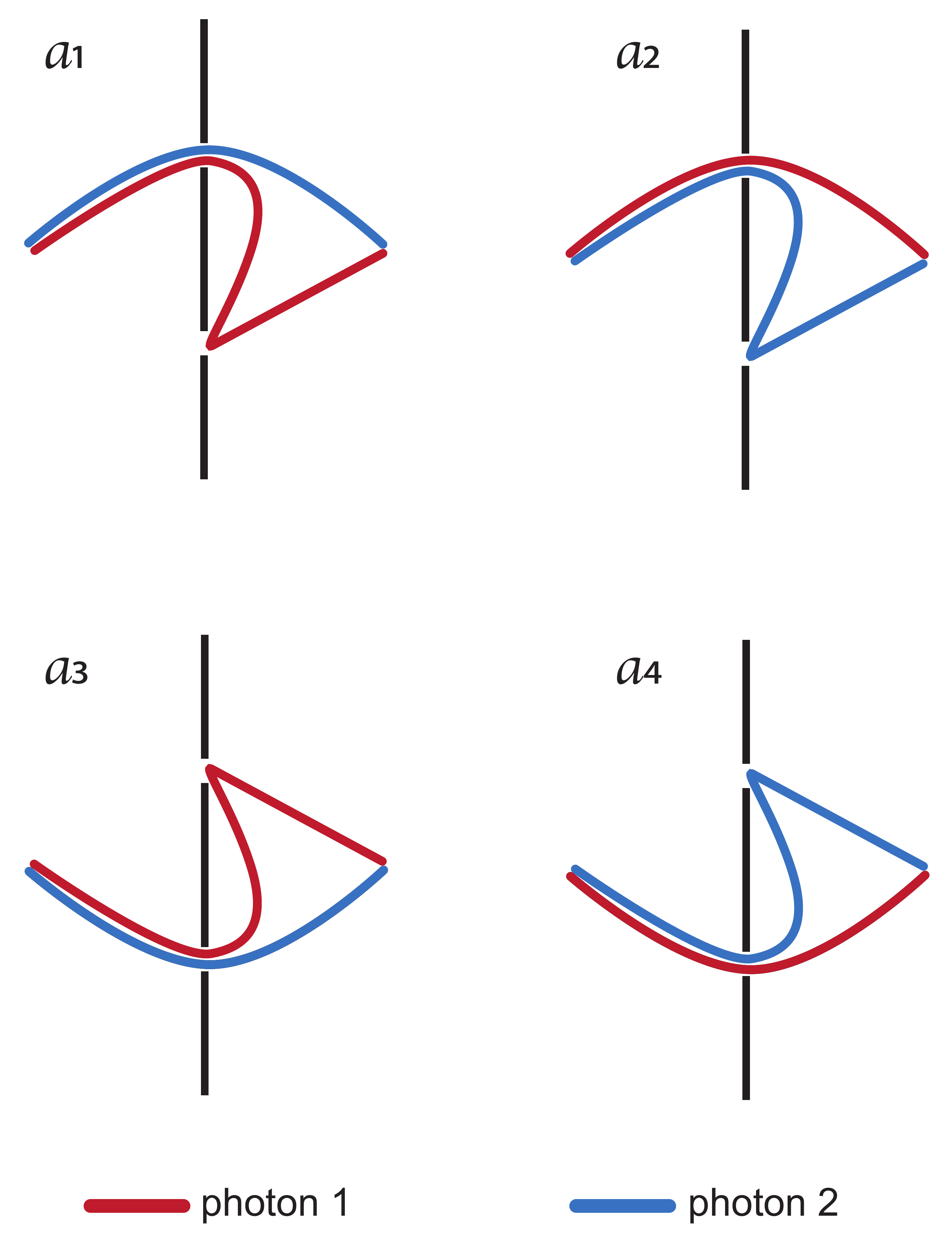}
\caption{The four contributions due to single-particle kink trajectories, when both photons go first through the same slit. These drawings represent same-position coincidence detection.}
\label{KinkA}
\end{figure}

\begin{figure}[h]
\centering
\includegraphics[width=5.0 cm]{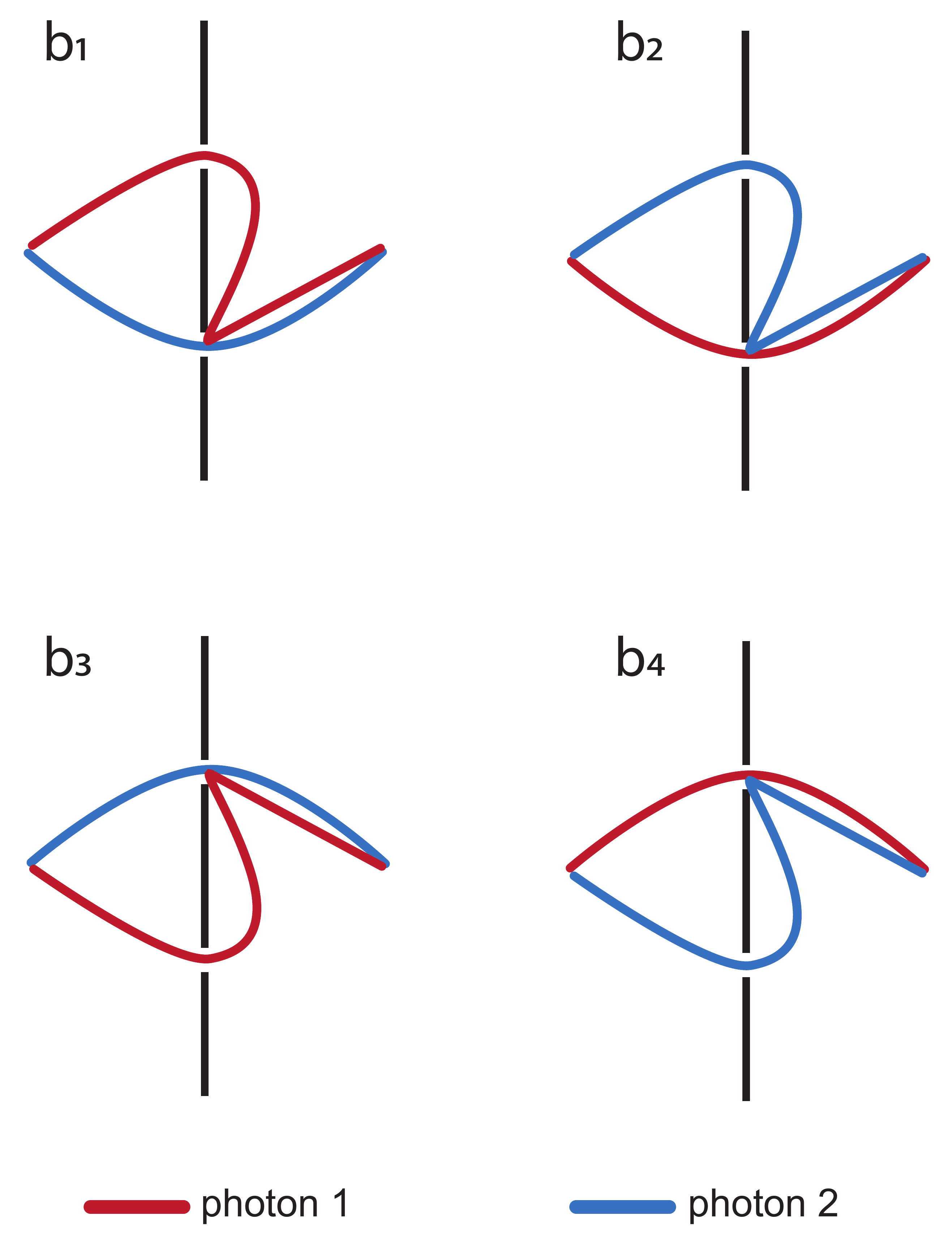}
\caption{The four contributions due to single-particle kink trajectories, when the photons go first through different slits. These drawings represent same-position coincidence detection.}
\label{KinkB}
\end{figure}

In a 2-slit interference experiment, we can have two types of non-classical trajectories, involving either kinks or loops around the slits as in figure \ref{KinkLoop}. For a 2-particle wavefunction one may include several possibilities namely one  particle performing a non-classical trajectory, while the other does a classical one, or both particles performing non-classical paths. For the propagation between the slits, we employ the propagator
\begin{align}\label{propagadorex}
G_\epsilon(x_i,t+\epsilon;x_0,t)&=\sqrt{\frac{1}{i\lambda c\epsilon}}\exp{\bigg[\frac{-\pi(x_i-x_0)^2}{i\lambda c\epsilon}\bigg]}.
\end{align}
An estimate for the inter-slit transit time  $\epsilon$   is given by $\epsilon=d/\Delta v_x$, where $\Delta v_x=\Delta p_x/\tilde{m}$,  $\tilde{m}=k/\hbar c$  \cite{OURPRA}) and $\Delta p_x=\sqrt{\langle p^2_x\rangle-\langle p_x\rangle^2}$ is the momentum uncertainty orthogonal to the propagation direction \cite{Paz2016,Geraldo2017,Vieira2019}. In the averages $\langle p^2_x\rangle$ and $\langle p_x\rangle^2$ we use the normalised wavefunction, including only classical trajectories, after the Gaussian-slit cropping. The simplest leading order contributions to the Sorkin parameter arise when one photon executes a kink and the other takes a classical trajectory, such as depicted in figures \ref{KinkA} and \ref{KinkB}. Loop contributions are less relevant by several orders of magnitude (as one particle goes three times through the slits), and so are the ones in which both particles perform-non-classical trajectories.

\begin{figure}[h]
\centering
\includegraphics[width=4 cm]{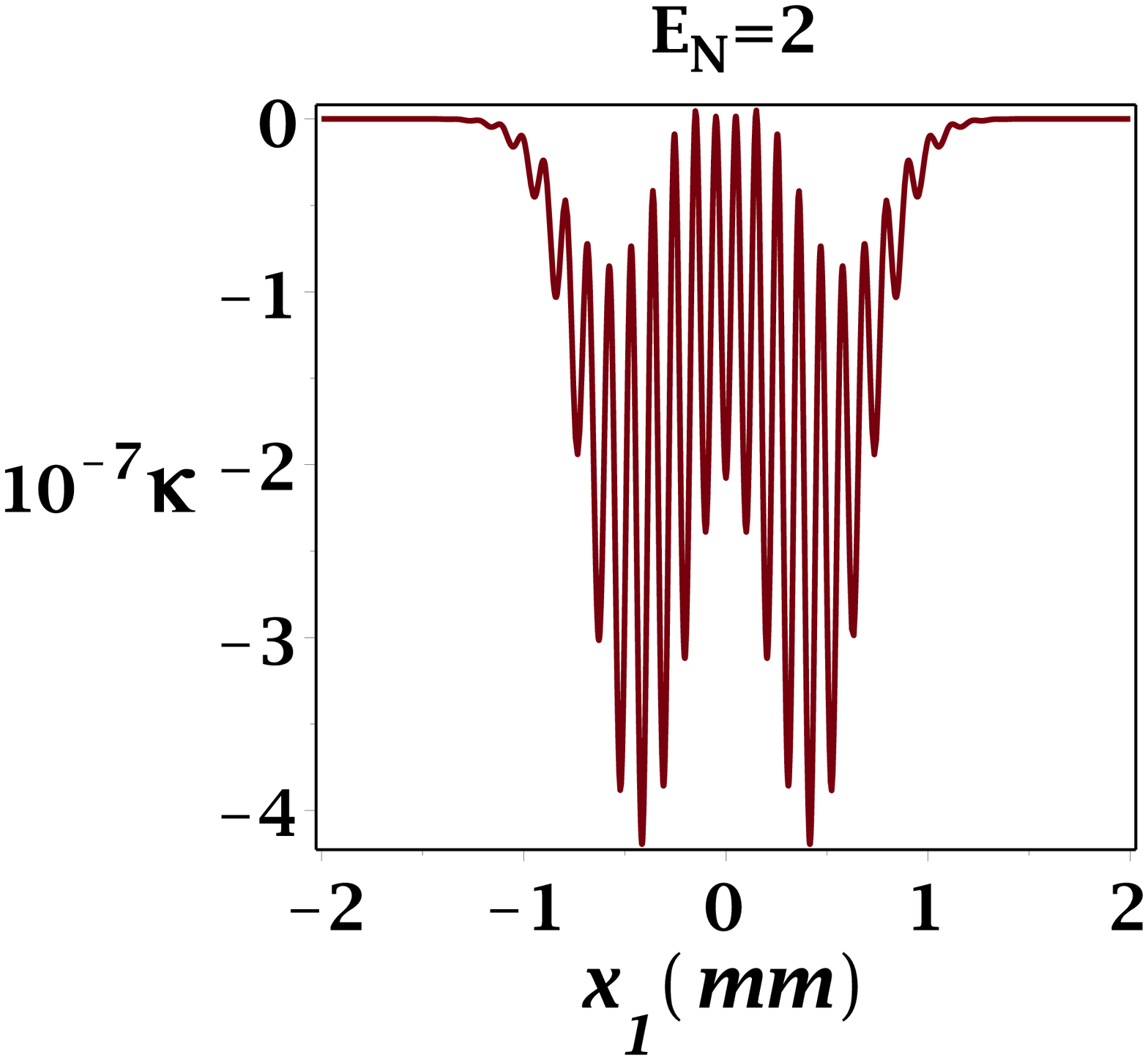}
\includegraphics[width=4 cm]{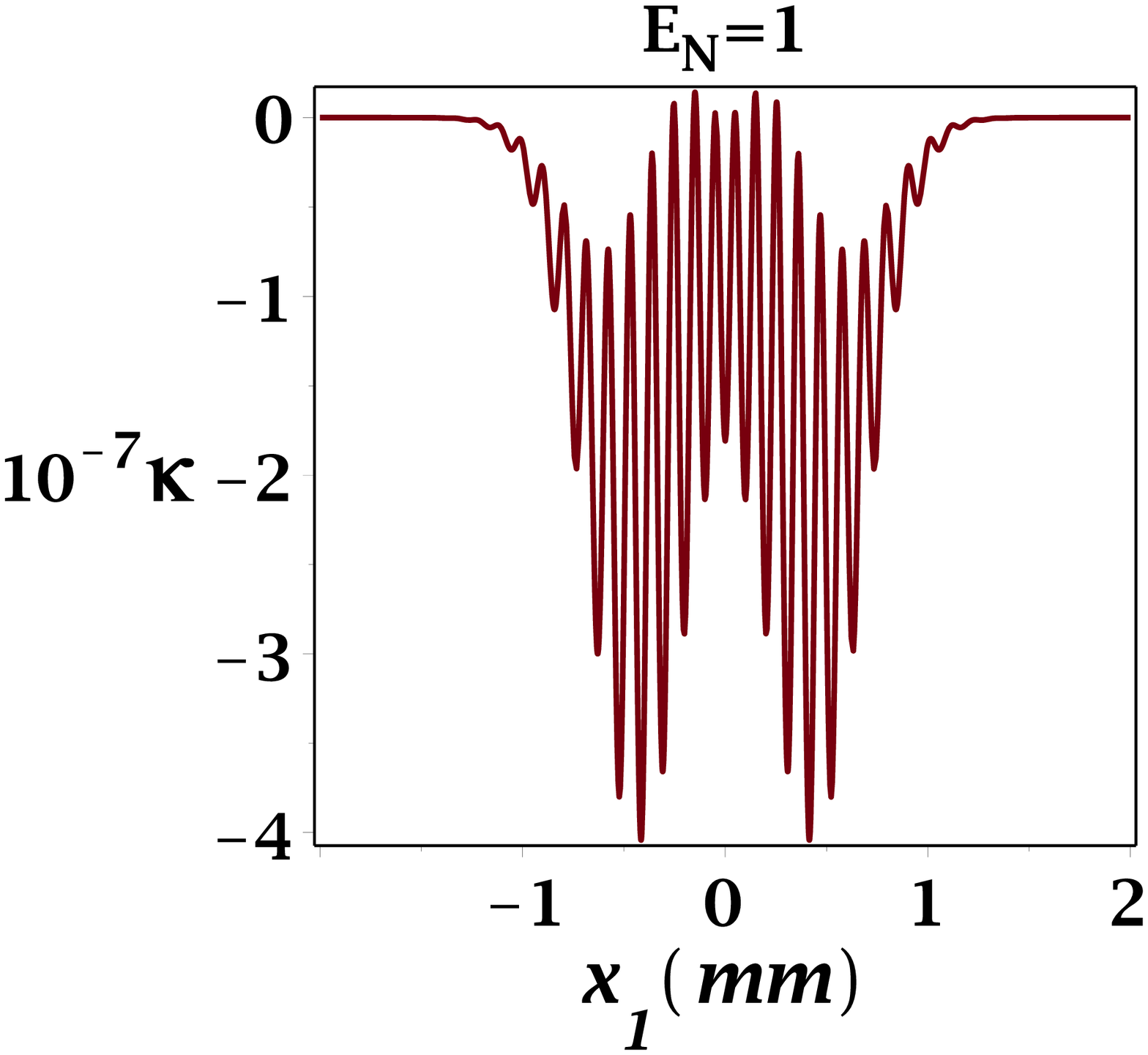}\\
\includegraphics[width=4 cm]{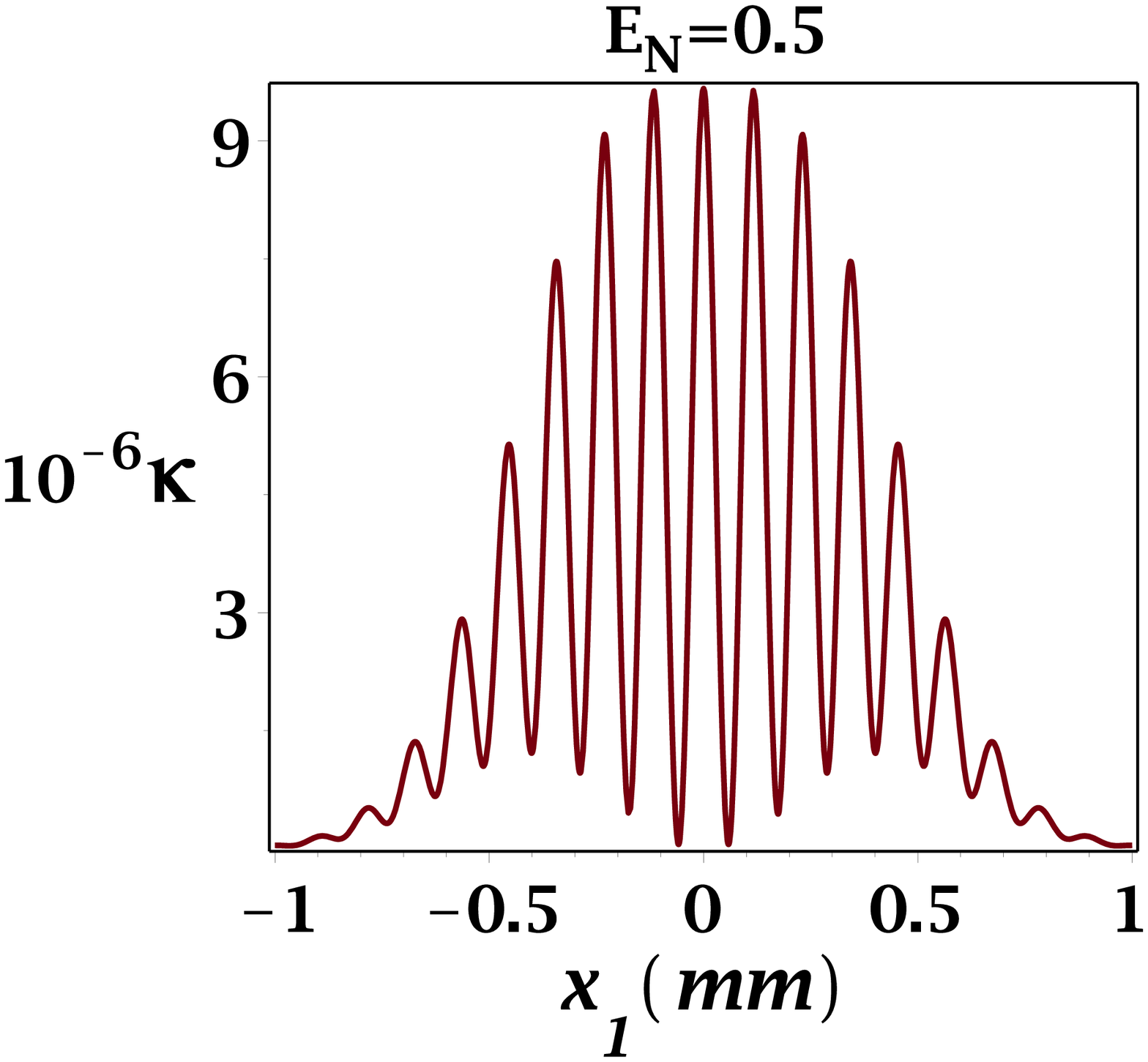}
\includegraphics[width=4 cm]{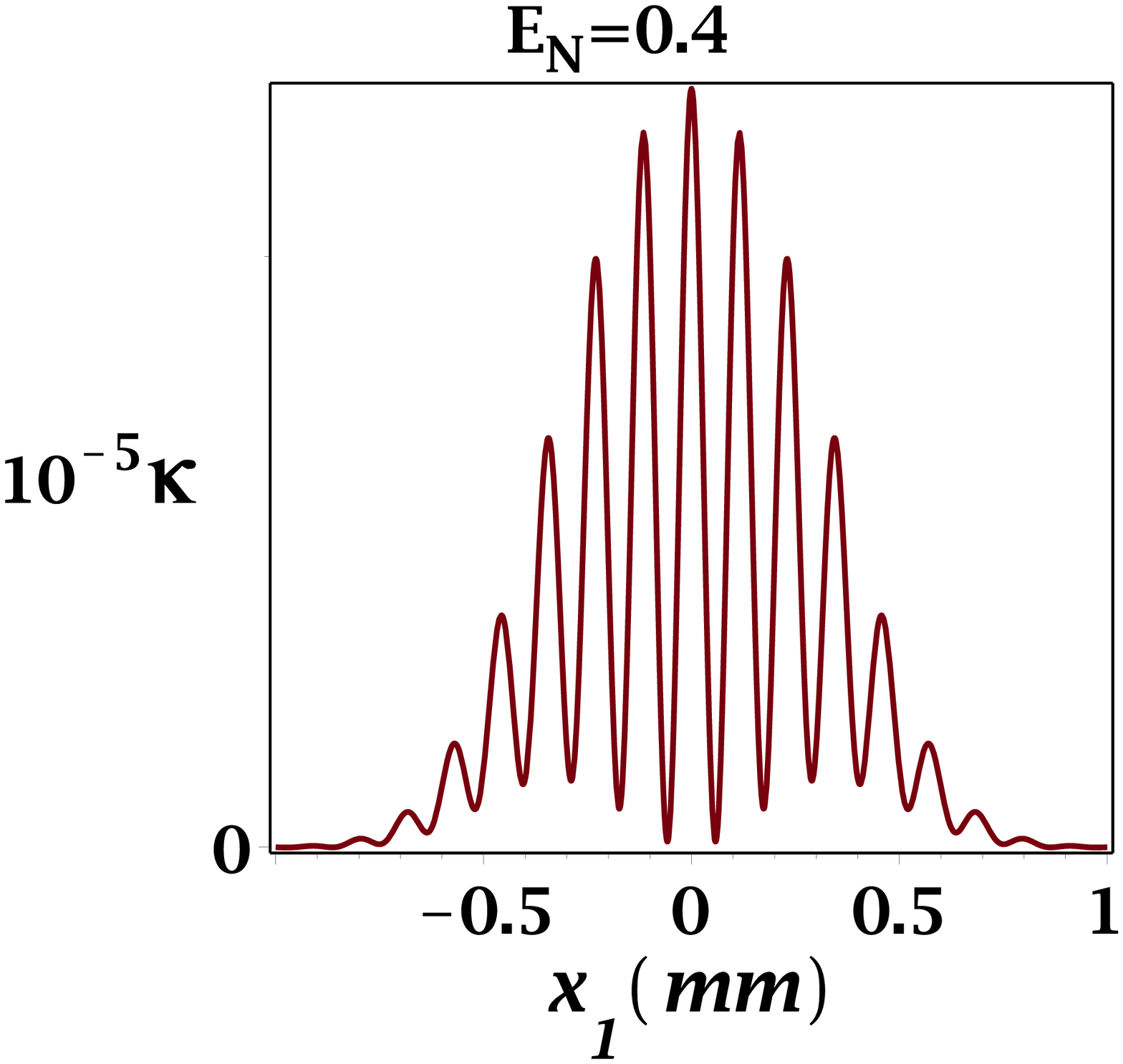}\\
\includegraphics[width=4 cm]{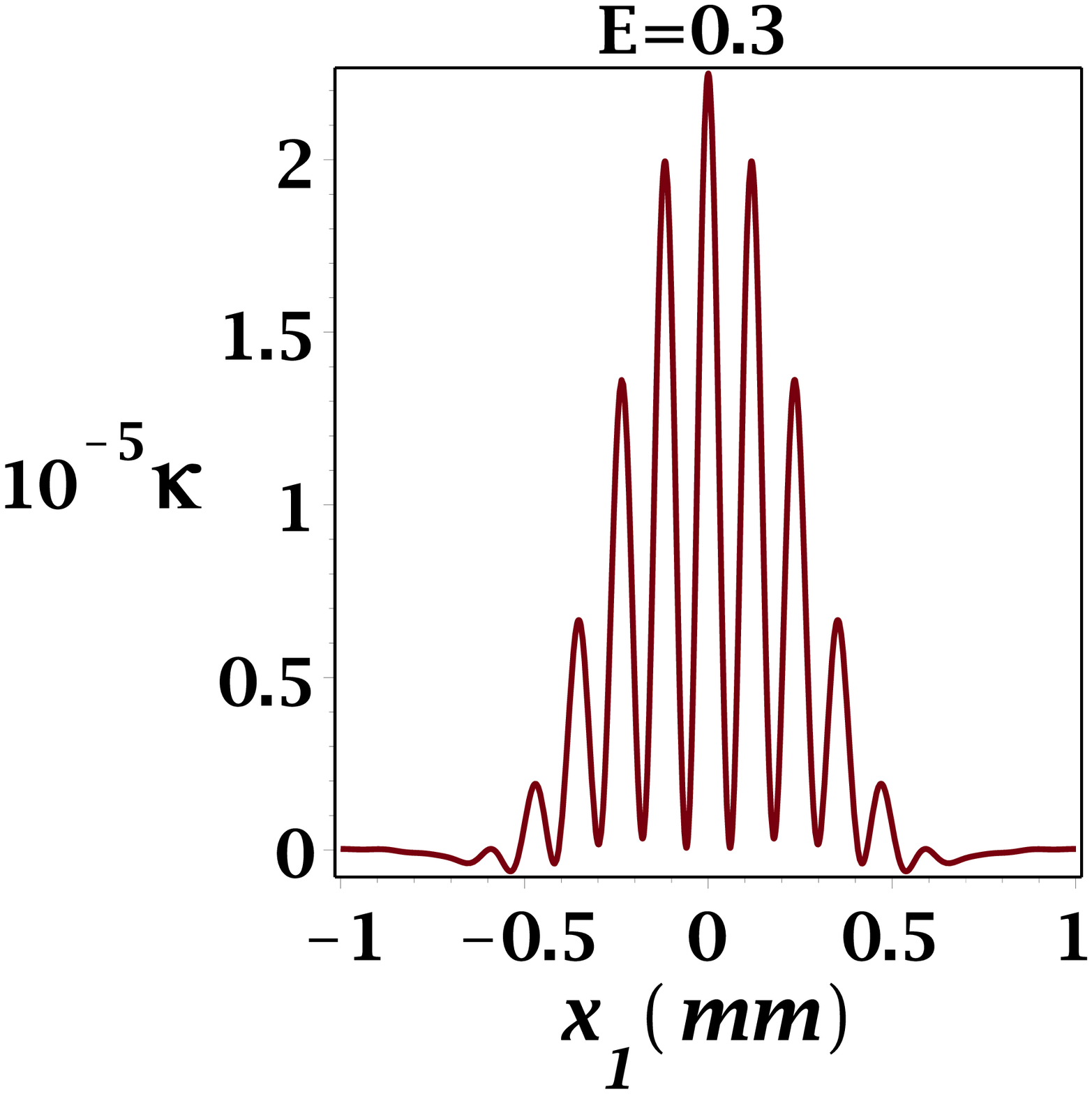}
\caption{Sorkin parameter $\kappa_{\text{nc}-a}$ for different values of logarithmic negativity $E_\mathcal{N}$. We used the set of parameters: $T=4\,$ps, $\tau=50\,$ps, $\sigma=11.4\,\mu$m, $\Omega=\sigma\times10^{E_{\mathcal{N}}}$, $\lambda=702\,$nm, $d=100\,\mu$m, $\beta=5\,\mu$m. In these plots we have set $x_2=0$ and swept over $x_1$. Observe that for $E_\mathcal{N}$ between $0.3$ and $0.4$ the Sorkin parameter increases about $2$ orders of magnitude if compared to $E_\mathcal{N} \approx 1$. The $x_1$ coordinates are plotted in millimeters. }
\label{Sorkin-a}
\end{figure}

The amplitude corresponding to photon 1 performing a kink (source $\rightarrow$ upper slit $\rightarrow$ lower slit $\rightarrow$ screen) while photon 2 takes a classical trajectory (source $\rightarrow$ upper slit $\rightarrow$ screen), as depicted in fugure \ref{KinkA}$(a_1)$, is obtained as
\bq
 &&\psi_{\text{nc}(a_1)}=\int G(x_1, T+\epsilon+\tau;x_1''',T+\epsilon)F_{d}(x_1''') \nonumber \\
 &&\times  G_\epsilon(x_1''', T+\epsilon;x_1'',T)F_{u}(x_1'') 
 G(x_2,T+\tau; x_2'', T)\nonumber \\ && \times F_{u}(x_2'')  G(x_1'',T;x_1', 0) G(x_2'',T;x_2', 0)\psi_0(x_1',x_2'),\label{nca1}
\eq
where the integral is over all primed variables $\{x_1',x_2',x_1'',x_2'',x_1'''\}$. Other possible non-classical paths for photons passing through the same slit are depicted  in figure \ref{KinkA}. They add up to
\begin{equation}
\Psi_{\text{nc}(a)}=\psi_{\text{nc}(a_1)}+\psi_{\text{nc}(a_2)}+\psi_{\text{nc}(a_3)}+\psi_{\text{nc}(a_4)},
\end{equation}
 whereas in figure \ref{KinkB}, the amplitudes for photons passing through different slits result in
\begin{equation}
\Psi_{\text{nc}(b)}=\psi_{\text{nc}(b_1)}+\psi_{\text{nc}(b_2)}+\psi_{\text{nc}(b_3)}+\psi_{\text{nc}(b_4)}.
\end{equation}
The weight of the contribution from the sets $a$ and $b$ is ruled by the degree of initial spatial correlation which, for the parameters specified in figure \ref{Rho}, is governed by 
$E_{\cal{N}}$ or ultimately by the ratio $\Omega/\sigma$. A lengthy but straightforward calculation for the analytical expressions of the contributions $\psi_{\text{nc}(a_i)}$ and $\psi_{\text{nc}(b_i)}$, $i=\{1,2,3,4\}$, can be easily computed using Maple$^\copyright$ \cite{Maple}:
\bq
\psi_{\text{nc}(a_i,b_i)} (x_1,x_2) &=& A_{\text{nc}(a_i,b_i)}\exp\Big[C_{\text{nc}(a_i,b_i)}(x_1,x_2) \nonumber \\ &+& i\,\, \alpha_{\text{nc}(a_i)}(x_1,x_2)\Big],
\eq
where the coefficients   $C_{\text{nc}(a_i,b_i)}(x_1,x_2)$ and $\alpha_{\text{nc}(a_i,b_i)}(x_1,x_2)$, omitting the $i$ index, have the general form
\begin{align}
C_{\text{nc}(a,b)}&\equiv \bar{c}_1 x_1^2 +\bar{c}_2 x_2^2+\bar{c}_3 x_1 x_2+\bar{c}_4 x_1+\bar{c}_5 x_2 + \bar{c}_6\nonumber\\
\alpha_{\text{nc}(a,b)}&\equiv \bar{a}_1 x_1^2 +\bar{a}_2 x_2^2+\bar{a}_3 x_1 x_2+\bar{a}_4 x_1+\bar{a}_5 x_2 +\bar{a}_6,\label{alphaCex}
\end{align}%
whose explicit expressions are found in appendix B. We shall vary the logarithmic negativity $E_\mathcal{N}$, which is related to  $\rho_x (t)$, to evaluate  the Sorkin parameter $\kappa_{\text{nc}(a,b)}$ due to the contributions given by $\psi_\text{nc(a,b)}$ that can be written as
\begin{equation}
\kappa_{\text{nc}(a,b)}=\frac{I_{\text{nc}(a,b)}-I_\text{c}}{I_{\text{nc}(a,b)}(0,0)},
\end{equation} 
where
\begin{equation}
I_{\text{nc}(a,b)}(x_1,x_2)=\vert\psi_{uu}+\psi_{ud}+\psi_{du}+\psi_{dd}+\psi_{\text{nc}(a,b)}\vert^2.\label{NCintensity}
\end{equation}
The intensity in Eq. (\ref{NCintensity}) depends on  $x_{1}$ and $x_{2}$ which means that the Sorkin parameter will depend on the measurement procedure. The coincidence measurements involving non-classical propagations takes a time interval  $\epsilon$  from slit to slit $\sim 1\,\, ps$. The time-resolution of current counters is of the order of $10^2\,\, ps$ \cite{PhotonCounter}, so the coincidence measurements are still inside the detectors' resolution. Our numerical analysis has shown that the Sorkin parameter for coincidence measurements such that $x_1 = x_2$ detection does not differ considerably from, say, $x_2=0$ and $x_1 = x$ on the screen.  Therefore, we will adopt the latter strategy. A few plots of $\kappa_{\text{nc}(a)}$ for different $E_\mathcal{N}$ values can be found in figure \ref{Sorkin-a}. The Sorkin parameter is a function of both $x_1$ and $x_2$, and we chose $x_1=x_2$ in the plots of figure \ref{Sorkin-a}. We remark that, in view of equation (\ref{sorkinDef2}), the Sorkin parameter assumes positive and negative values in the $x_1-x_2$ plane, since its integral over it should be zero for normalized intensities.

\begin{figure}[h]
\centering
\includegraphics[width=7cm]{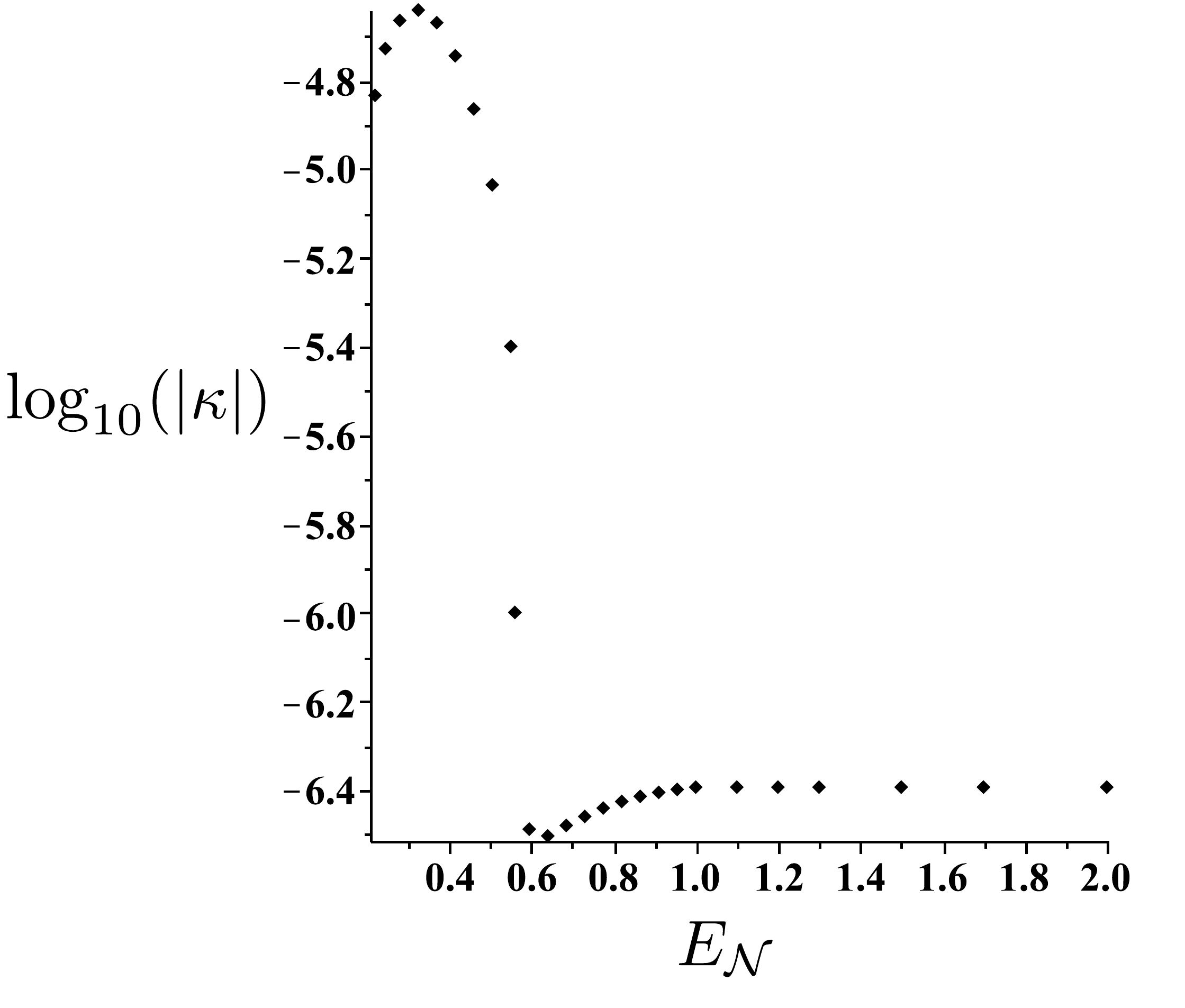}
\caption{Double-slit: The behaviour of the maximum value of the Sorkin parameter as a function of $E_\mathcal{N}$ which shows that it has a maximum value $\kappa \approx 10^{-5}$ for $0.3 \le E_\mathcal{N} \le 0.4 $ before stabilising at $10^{-7}$ for $E_\mathcal{N} > 1$.}
\label{Sorkin-aa}
\end{figure}

The Sorkin parameters shown in figure \ref{Sorkin-a} have values that are about one order of magnitude lower than the ones found  in \cite{Sinha15} (in which a 3-slit setup for single photons was used) in the range $E_\mathcal{N} \apprge 1$.  Remarkably, in the range $E_\mathcal{N}=[0.2, 2]$  the Sorkin parameter increases by about $2$ orders of magnitude for $ 0.3 \apprle E_\mathcal{N} \apprle 0.4$ ($\kappa \approx 10^{-5}$), in comparison to $E_\mathcal{N}\apprge 1$. Qualitatively, large values of $E_\mathcal{N}\apprge1$ do not mean a larger Sorkin parameter, since photons are unlikely to separate in a nonclassical trajectory during the inter-slit transit time, due to their spatial correlation. There is, however, an optimal region $ 0.3 \apprle E_\mathcal{N} \apprle 0.4$ that yields an increase to the Sorkin parameter, because the nonclassical trajectories are not as suppressed as for large negativities $E_\mathcal{N}$. On the other hand, for $E_\mathcal{N}\apprle 0.2$, we observed that the Sorkin parameter becomes negligible since the biphotons are unlikely to diffract through the same slit. Figure \ref{Sorkin-aa} shows the plot of $\log_{10}(|\kappa|)$ against $E_\mathcal{N}$ for the double slit parameters
specified in figure \ref{Sorkin-a}.

\begin{figure}[h]
	\centering
	\includegraphics[width=6.5cm]{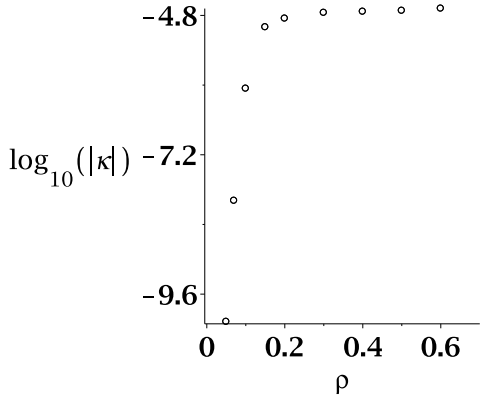}
	\caption{Double-slit: the behaviour of the maximum value of the Sorkin parameter as a function of $\rho_x$. We used the set of parameters: $\tau=50\,$ps, $\sigma=11.4\,\mu$m, $\Omega=\sigma\times10^{E_{\mathcal{N}}}$, $E_{\mathcal{N}}=0.4$, $\lambda=702\,$nm, $d=100\,\mu$m, $\beta=5\,\mu$m.}
	\label{Sorkin-aaa}
\end{figure} 

Moreover the contributions from the trajectories in figure \ref{KinkB} to the Sorkin parameter $\kappa_{\text{nc}(b)}$ are negligible compared to $\kappa_{\text{nc} (a)}$. This is a consequence of the short transit time between the SPDC crystal and the slits, which favours biphotons going first through the same slit. Accordingly, by increasing the transit time between source and slits, the contributions of types $\kappa_{\text{nc} (a)}$ and $\kappa_{\text{nc} (b)}$ become comparable. Numerical evaluations have shown that, by using the same set of parameters from figure \ref{Sorkin-a},  $E_\mathcal{N}=\{2,1,0.5,0.4,0.3\}$ one obtains, respectively $\kappa_{\text{nc}(b)}\sim\{10^{-22},10^{-22},10^{-17},10^{-15},10^{-13}\}$.

 The spatial correlations as given by the Pearson's value at the slits, $\rho_x(T)$, do not determine by themselves the value of the Sorkin parameter. The logarithmic negativity $E_{\mathcal{N}} (\sigma, \Omega)$, which is constant up to the grating, is related to $\rho_{x} (t)$ through equation (\ref{pearson}). For the double-slit setup and the exotic paths of the configuration $\kappa_{\text{nc} (a)}$, we can study how the Sorkin parameter varies with $\rho_{x} (T)$, $T$ being the typical flight time from the source to the grating. In figure \ref{Sorkin-aaa} we plot the (logarithm of) $\kappa$ as a function of the Pearson's value after choosing $E_{\mathcal{N}} = 0.4$. We can see that the highest Sorkin parameter ($\approx 10^{-4.8}$), is achieved for $\rho_x (T) \ge 0.2$.

\begin{figure}[h]
\centering
\includegraphics[width=6.0 cm]{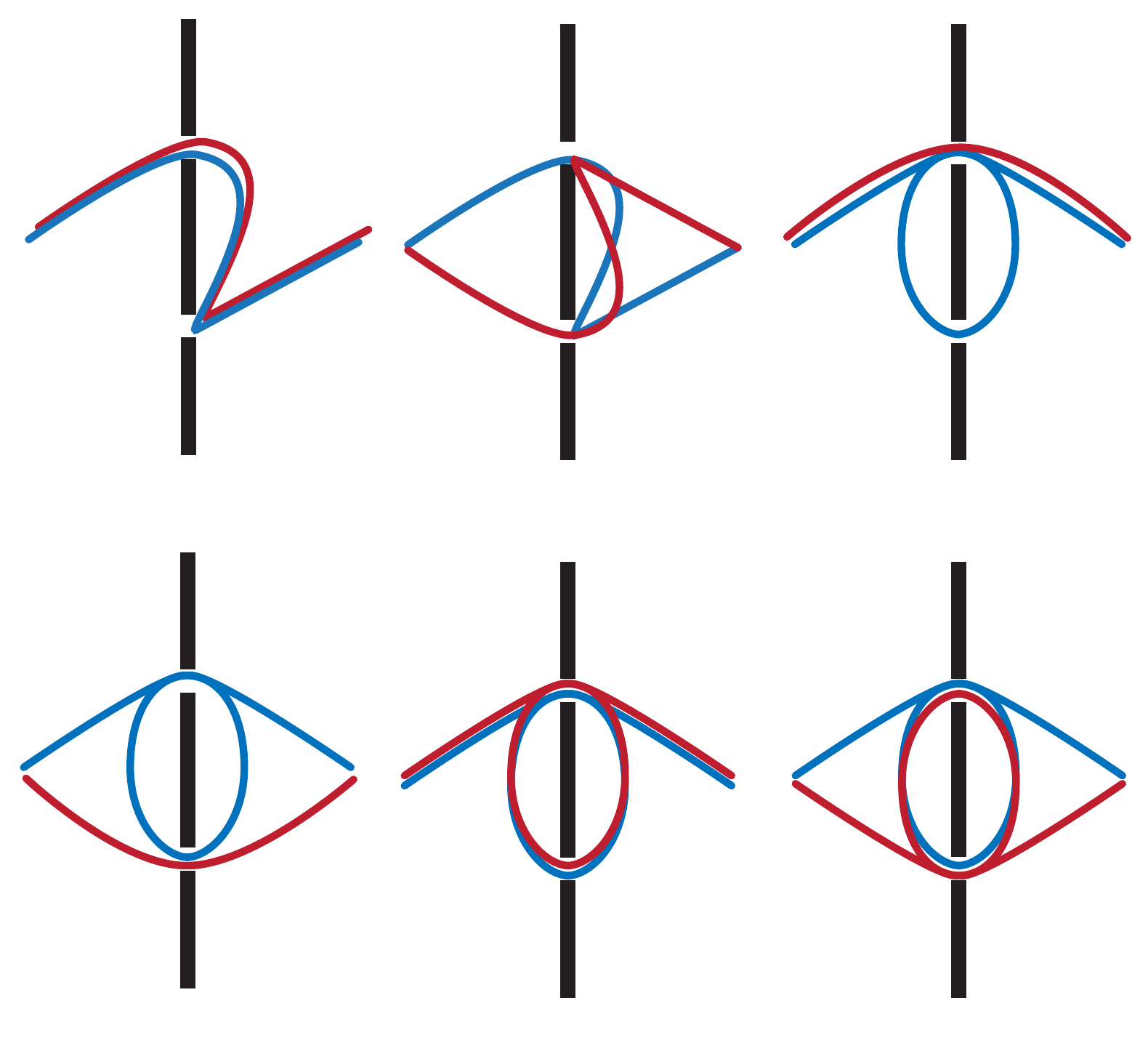}
\caption{Other types of non-classical contributions to the wave function at the screen. Their contribution to the Sorkin parameter is at least around 8 orders of magnitude lower than the one arising from paths in figure \ref{KinkA}, for the same set of parameters used in figure \ref{Sorkin-a} besides $E_\mathcal{N}=2$. For this value of $E_\mathcal{N}$, the top left contribution is dominant, and yields a Sorkin parameter of about $10^{-15}$. The top-right looped trajectory yields a Sorkin parameter of order $10^{-16}$ for the same set experimental values, placing looped trajectories in lower relevance as compared to kink ones.}
\label{ExtraPaths}
\end{figure}

Finally one may construct other sets of non-classical trajectories as shown in figure \ref{ExtraPaths}. Their contributions to the Sorkin parameter depends on the value of $E_\mathcal{N}$ chosen. For a typical value of the ratio $\Omega/\sigma \approx 100$ \cite{Schneeloch2016}, yielding $E_\mathcal{N}=2$, they are at least about 8 orders of magnitude below the dominant contribution from the paths in figure \ref{KinkA}. A reasonable rule of thumb is: the more slits the photons go through, the lower their contribution.

\subsection{Sorkin parameter for the biphoton: triple-slit}

In order the evaluate the Sorkin parameter for light waves without any enhancement mechanism as in \cite{Boyd16}, triple-slit photon interference was described in \cite{Sinha14,Sinha15,FDTD}.

\begin{figure}[h]
	\centering
	\includegraphics[width=5 cm]{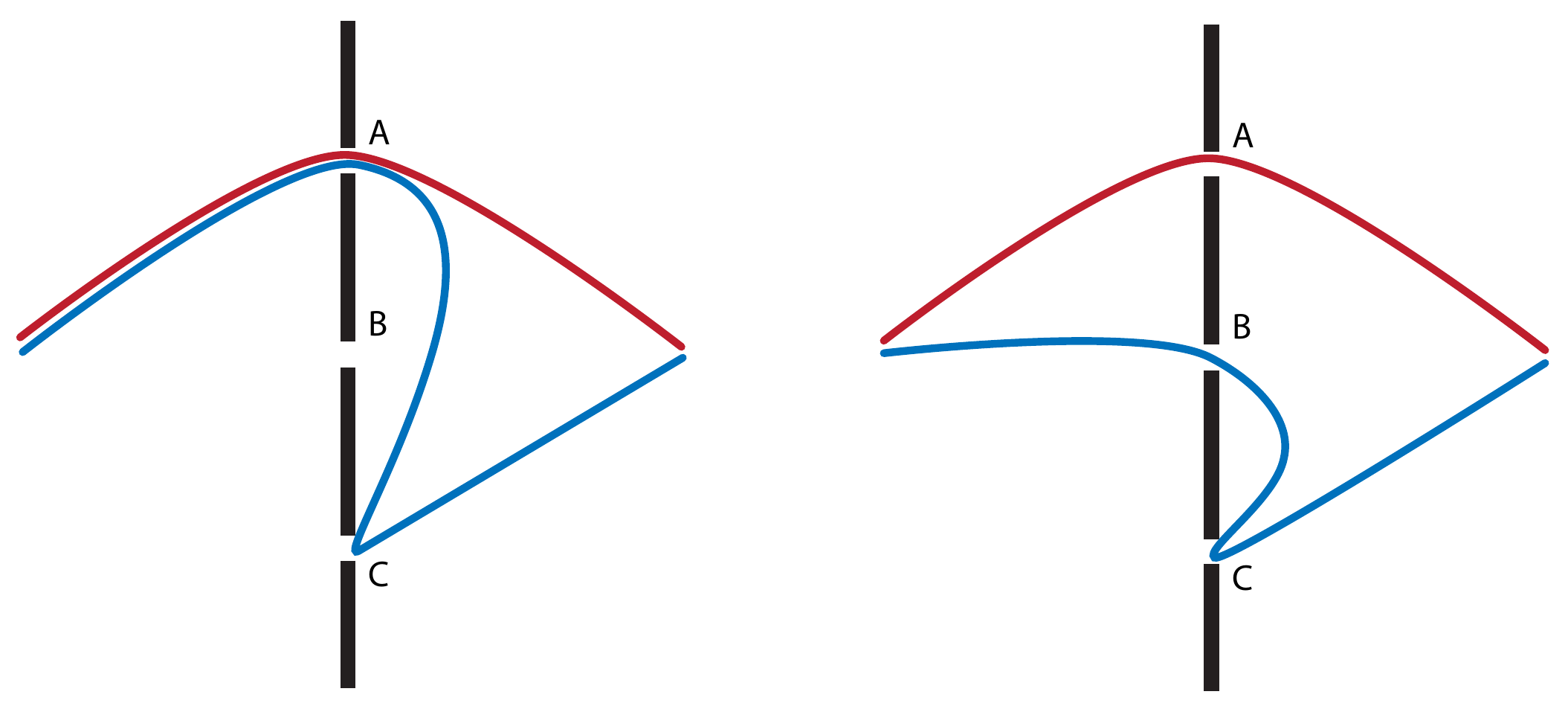}
	\caption{Typical leading order contribution to non-classical paths of biphotons}
	\label{3slitsBP}
\end{figure}

\begin{figure}[h]
	\centering
	\includegraphics[width=8 cm]{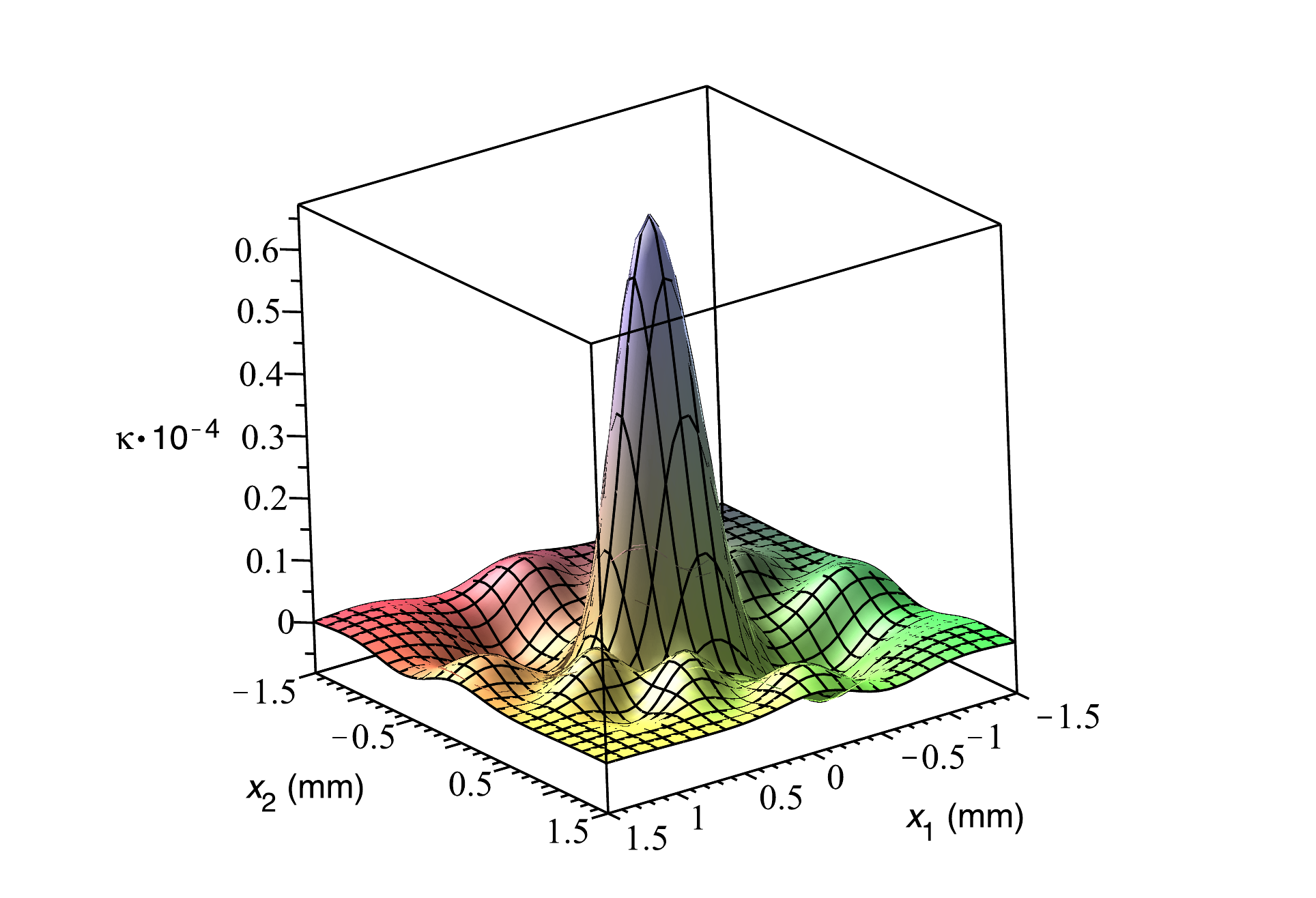}
	\caption{Sorkin parameter for biphoton 3-slit interference. We have adopted $E_{\cal{N}} = 2.0$, $T \approx 600 \,\, ps$ for which $\rho_{x}(600\,\, ps) = 0.026$, $\sigma=11.4 \,\, \mu m$, $\lambda = 810\,\, nm$, $\beta = 30\,\, \mu m$, $d= 100\,\, \mu m$, $\tau = T$.}
	\label{3slitsBPS3D}
\end{figure}

 The typical leading order contributions to non-classical paths are depicted in figure \ref{3slitsBP}. We adopt a set of parameters similar to those chosen in \cite{Sinha14} as shown in figure \ref{3slitsBPS3D}. In addition, the value of $\kappa$ is insensitive to whether coincidence measurements  performed at the same point $x_1 = x_2 = x$ or  one detector is fixed at say $x_2 = 0$, and $x_1 =x $ is an arbitrary point at the detection screen. In this optimised setup, the resulting Sorkin parameter is approximately $10^{-5}$. For  $E_{\cal{N}} = 2.0$ the relevant non-classical contributions come mainly from biphotons that pass through the same slit (paths like those on the left of figure \ref{3slitsBP}). The Sorkin parameter is defined and evaluated in a similar fashion as for the double-slit case, 
\be
\kappa_{\text{nc}}(x_1,x_2) = \frac{I_\text{nc}(x_1,x_2)-I_\text{c}(x_1,x_2)}{I_\text{nc}(0,0)}.
\ee

Figure \ref{3slitsBPS3D} illustrates the profile of the Sorkin parameter for coincidence  measurements for arbitrary $x_1$ and $x_2$. 

\begin{figure}[h]
	\centering
	\includegraphics[width=8cm]{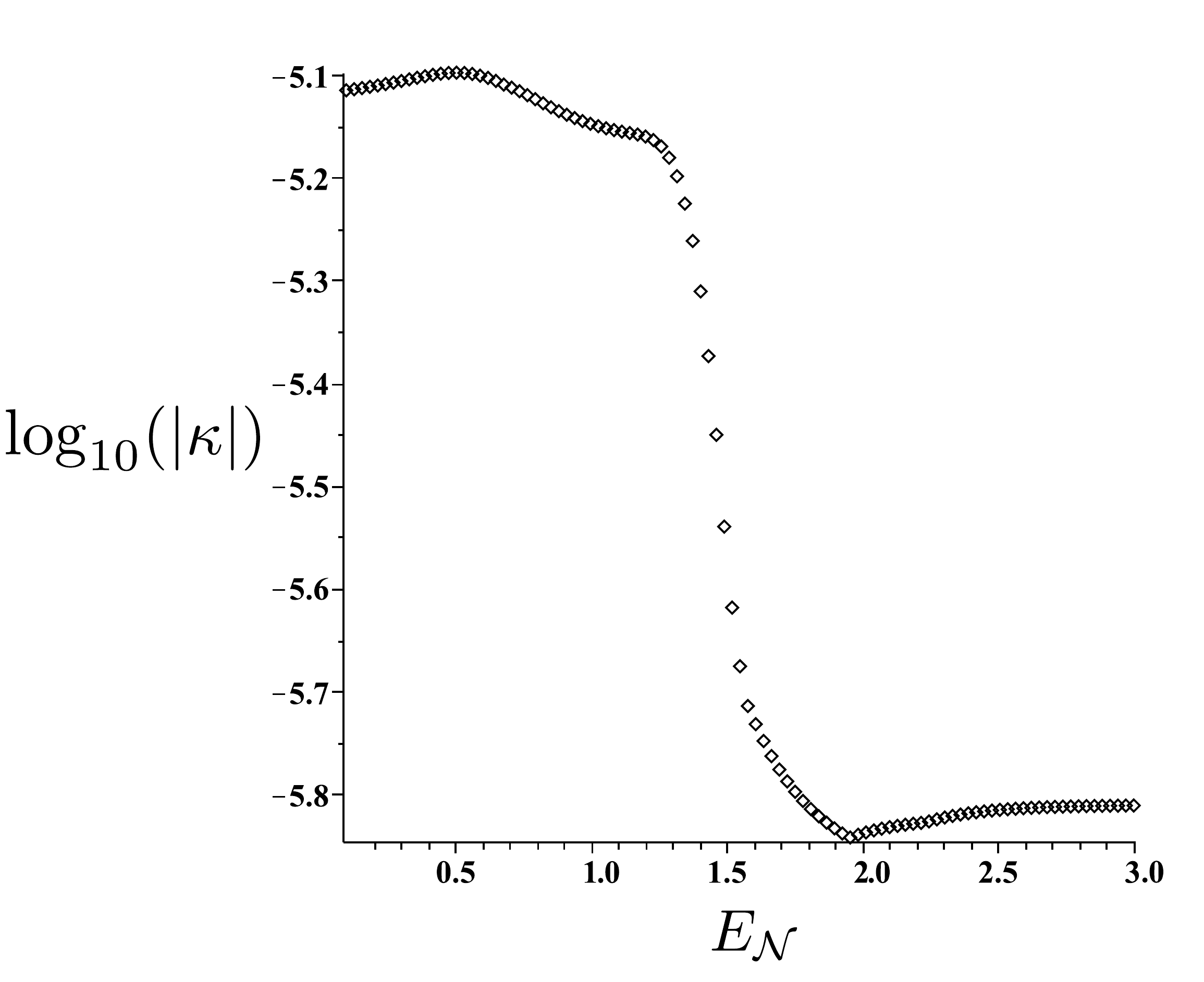}
	\caption{Triple-Slit: Maximal Sorkin parameter for biphoton 3-slit interference as a function of $E_{\cal{N}}$. We have adopted $T \approx 60 \,\, ps$,  $\sigma=11.4 \,\, \mu m$, $\lambda = 810\,\, nm$, $\beta = 10\,\, \mu m$, $d = 250\,\, \mu m$, $\tau = 270 \,\, ps$. }
	\label{SN-TS}
\end{figure}

Curiously and unlike the double-slit case shown in the previous section, the Sorkin parameter for coincidence measurements in a biphoton  triple-slit setup is not as sensitive to the logarithmic negativity. There are no significant changes in the order of magnitude for the set of parameters we have chosen. However for a different set of parameters as shown in figure \ref{SN-TS} it displays a similar behaviour as the double-slit for the maximum Sorkin parameter as a function of the negativity. The maximum value of $\kappa$ is still around $10^{-5}$  for $E_{\cal{N}} = 0.5$.

\subsection{Sorkin parameter for a massive particle}
\subsubsection{triple-slit setup}

In this section we will evaluate the Sorkin parameter for electron waves in a three-slit setup just as in \cite{Sinha14} in order to assess the efficiency of our simplified model. 

The effective propagator for a free particle of mass $m$ reads \cite{Feynman}
\begin{align}\label{mpropagador}
G(x,t;x_0,t_0)&=
\sqrt{\frac{m}{2\pi i \hbar \left(t-t_0\right)}}\exp{\bigg[\frac{-m(x-x_0)^2}{2i\hbar\left(t-t_0\right)}\bigg]},
\end{align}
was employed in \cite{Paz2016,Geraldo2017,Vieira2019} to study Gouy phases, matter wave interferometry and exotic paths contributions to the Sorkin parameter.

\begin{figure}[h]
\centering
\includegraphics[width=4 cm]{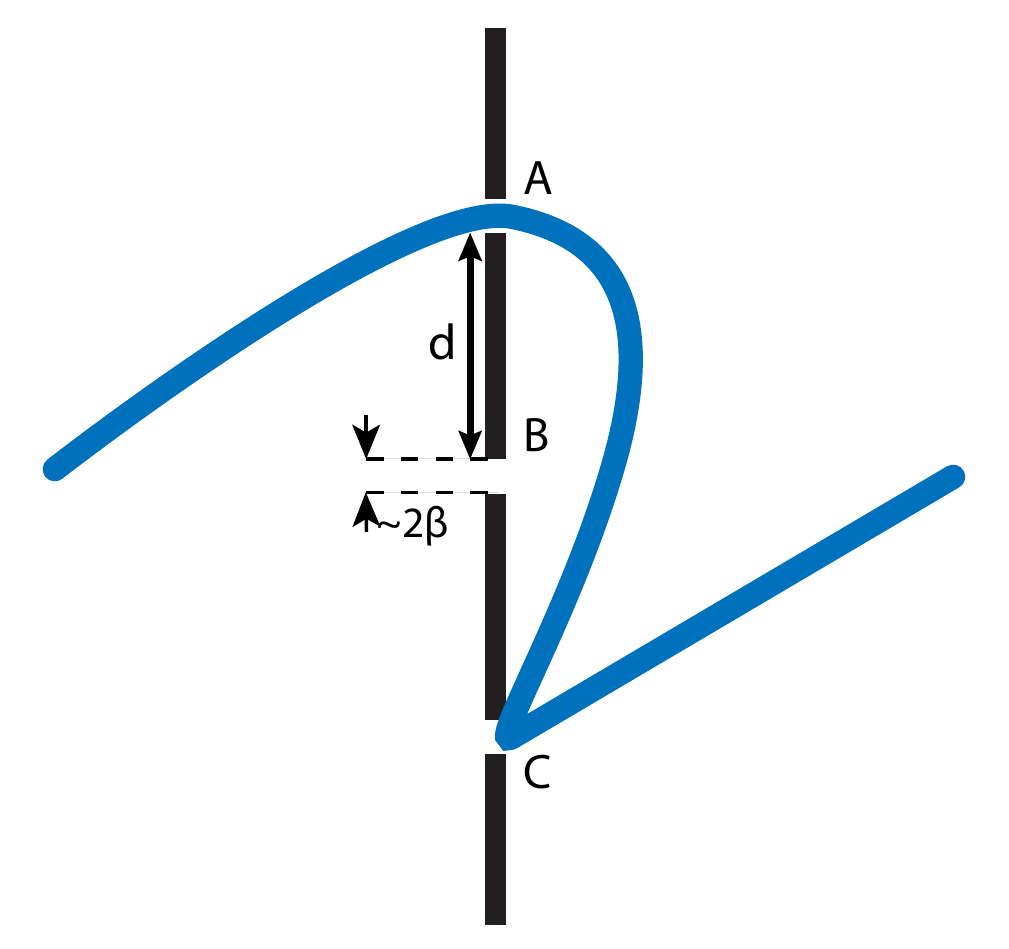}
\caption{An example of kink type non-classical trajectory in a 3-slit Young experiment.}
\label{3slits}
\end{figure}

Consider kink-like trajectories such as the one shown in figure \ref{3slits} in which the slits are labeled A, B and C.
A classical path amplitude contribution at the screen corresponding to, say, the particle going through the slit $j$ reads
\begin{equation}
 \psi_{j}(x)=\int_{x'', x'} G(x,T+\tau;x'',T)F_j(x'') G(x'',T;x',0)\psi_0(x'),
 \label{mpsi-classical}
\end{equation}
where
\begin{equation}
 \psi_{0}({x'})=\frac{1}{\sqrt{\sigma_0\sqrt{\pi}}}\exp\left(-\frac{x'^2}{2\sigma^2_0}\right)
  \label{psi_0}
\end{equation}
is the initial Gaussian wavepacket, in which the standard deviation $\sigma_0$ is related to the collimator size. The window functions $F_j(x'')$ that modulate the slit apertures can be written as 
\be
 F_{A,C}(x'') =e^{-\frac{(x'' \mp d)^2}{2\beta^2}} \,\, \text{and} \,\, 
 F_B(x'') = e^{-\frac{(x'')^2}{2\beta^2}}. 
\ee
The non-classical trajectory contributions are represented by $\psi_{jl}$, meaning the particle goes through slit $j$, then to slit $l$, then to the screen. They are evaluated as the following
\begin{align}
 \psi_{jl}(x)=&\int_{x''',x'', x'} G(x,T+\delta+\tau;x''',T+\delta)\nonumber\\
 &F_l(x''')G(x''',T+\delta;x'',T)F_j(x'') \nonumber\\
 &G(x'',T;x',0)\psi_0(x'),
 \label{mpsi-non-classical}
\end{align}
where the parameter $\delta$ corresponds to the inter-slit transit time; for slits separated by $d$ ($2d$), it evaluates to $\epsilon$ ($2\epsilon$). The parameter $\epsilon$ is evaluated using $\epsilon=d/\Delta v_x$, where $\Delta v_x=\Delta p_x/m$, in which $\Delta p_x=\sqrt{\langle p^2_x\rangle-\langle p_x\rangle^2}$ is the momentum variance orthogonal to the propagation direction.

\begin{figure}[h]
\centering
\includegraphics[width=6 cm]{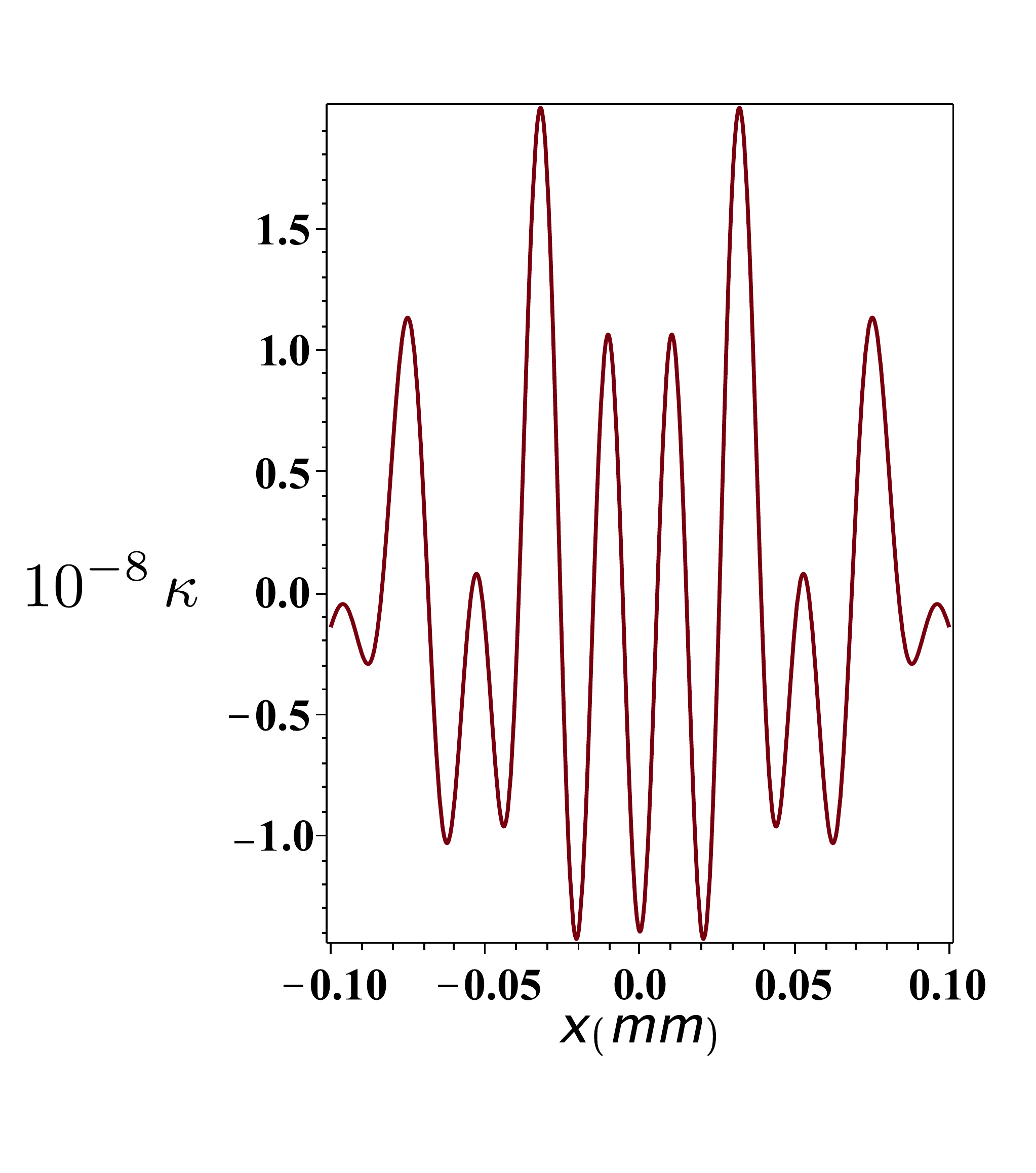}
\caption{The Sorkin parameter due to trajectories of type depicted in figure \ref{3slits} in a 3-slit setup. We have chosen to use an electron with de Broglie wavelength $50\,$pm, $d=272\,$nm, $\beta=31\,$nm, and $\sigma_0=62\,$nm. The source-to-slit distance  is $24\,$cm, and the slits-to-screen one is $30.5\,$cm. The variable $x$ is plotted in mm.}
\label{sorkin3slit}
\end{figure}

We evaluate the Sorkin parameter following  \cite{Sinha15}, as discussed in the introduction. It reads:
 \begin{equation}
\kappa=\frac{\Delta I}{I_0},
\end{equation}
where, to first order in the path contributions
\begin{align}
\Delta I\approx 2\Re[\psi^*_A(\psi_{BC}&+\psi_{CB})+\psi^*_B(\psi_{AC}+\psi_{CA})\nonumber\\
&+\psi^*_C(\psi_{AB}+\psi_{BA})],
\end{align}
and $I_0$ is the total intensity at the central peak. A plot of the Sorkin parameter for the parameters used for electron waves in Ref. \cite{Sinha15} is found in figure \ref{sorkin3slit}. The order of magnitude obtained with our effective description agrees with the one obtained in \cite{Sinha15}, which validates our effective description.

\subsubsection{double-slit setup}

It is instructive to rank the contributions to the Sorkin parameter in our framework for the interference one massive particle in a 2-slit experiment arising from: (a) non-classical kink paths (b) non-classical looped path trajectories, and (c) relativistic corrections to the propagator. Because the contributions from non-classical paths are very small, it is natural to ask how they compare to relativistic corrections, even for a small average velocities of particles (as compared to the speed of light) in the source beam.

Now let us proceed to rank the contributions of non-classical paths (kinks or loops) for a double-slit setup as well as compare to relativistic corrections to the propagators. We shall use matter waves for neutrons and electrons.

Relativistic corrections can be implemented via a simple modification in the propagator as discussed in \cite{pad}
\begin{align}
G_\text{rel}(x,t;x_0,t_0)=&G(x,t;x_0,t_0)\bigg[1-\frac{3(x-x_0)^2}{4c^2(t_E-t_{E_0})^2}\nonumber\\
&+\frac{m(x-x_0)^4}{8 \hbar c^2(t_E-t_{E_0})^3}+\mathcal{O}\bigg(\frac{1}{c^4}\bigg)\bigg],
\label{Krel}
\end{align}
where $G(x,t;x_0,t_0)$ is given by Eq. (\ref{mpropagador}) and $t_E$ stands for the Euclidian time, that is, $t_E=it$.
Because the relativistic corrections are small, we will use them on the classical trajectories only. Hence, we have three distinct scenarios: (a) non-classical kink-type trajectories, (b) non-classical looped  trajectories, and (c) relativistic corrections to the propagator.

The non-classical kink-like trajectories are found in the same way as in Eq. (\ref{mpsi-non-classical}). The loop contribution corresponding to a path such as in figure \ref{KinkLoop} (b) is evaluated as%
\begin{align}
 \psi_{\text{loop}-jk}(x)=&\int G(x,T+2\epsilon+\tau;x'''',T+2\epsilon)F_j(x'''')\nonumber\\
 &\times G(x'''',T+2\epsilon;x''',T+\epsilon)F_k(x''')\nonumber\\
 &\times G(x''',T+\epsilon;x'',T)F_j(x'') \nonumber\\
 &\times G(x'',T;x',0)\psi_0(x'),
 \label{mpsi-loop}
\end{align}
which should be read as ``the particle goes first through slit $j$, then loops through slit $k$, and propagates from slit $j$ to the screen''. The integrals are carried out over all primed coordinates $\{x',x'',x''',x''''\}$, and their analytical forms are shown in appendix C. The relativistic corrections, on the other hand, are implemented by substituting the propagator in Eq. (\ref{mpsi-classical}) by its corrected version in Eq. (\ref{Krel}).

\begin{figure}[h]
\centering
\includegraphics[width=6 cm]{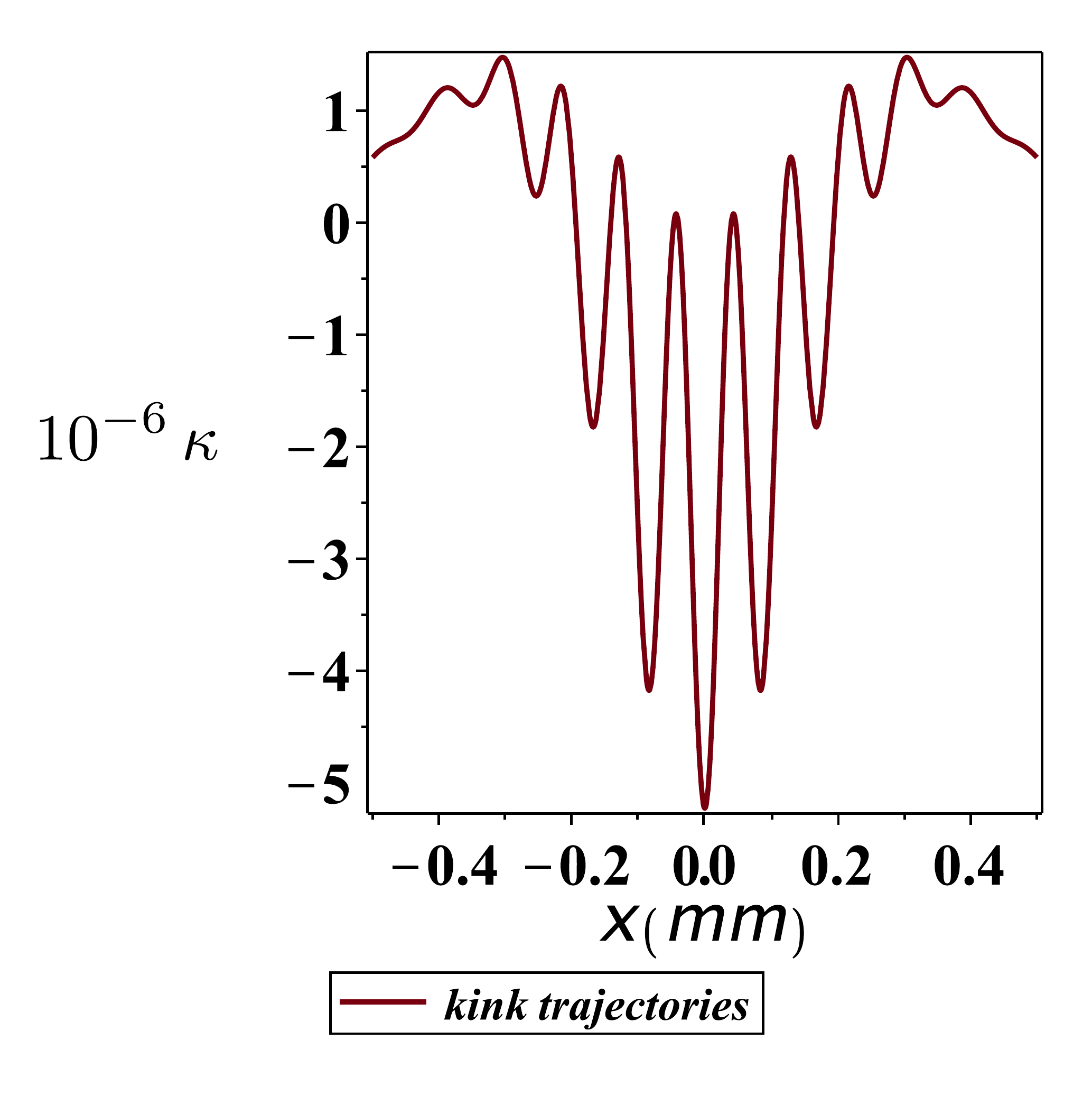}
\includegraphics[width=6 cm]{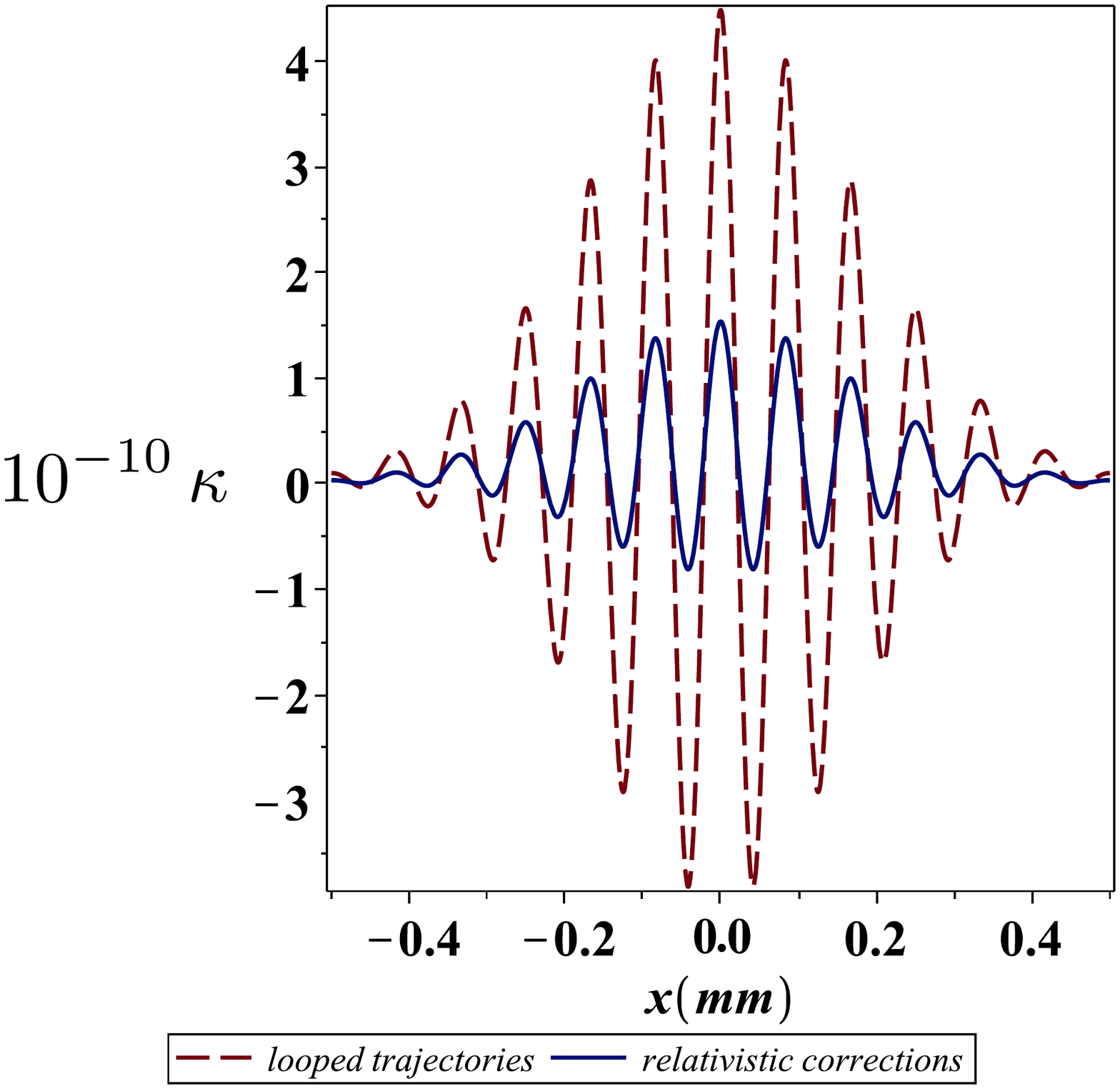}
\caption{The Sorkin parameter due to 3 different contributions for a neutron going through a double-slit. The neutron has Broglie wavelength $2\,$nm, and the remaining parameters are $d=125\,\mu$m, $\beta=7\,\mu$m, $\sigma_0=7\,\mu$m, and $t=\tau=26.4\,$ms. For these parameters, the inter-slit transit time $\epsilon$ is $18$ ms. The variable $x$ is plotted in mm. In this case it is clear that the Sorkin parameter is generated mainly by kink-like trajectories, as the other contributions are about 4 orders of magnitude lower.} 
\label{mrank}
\end{figure}
\begin{figure}[h]
\centering
\includegraphics[width=6 cm]{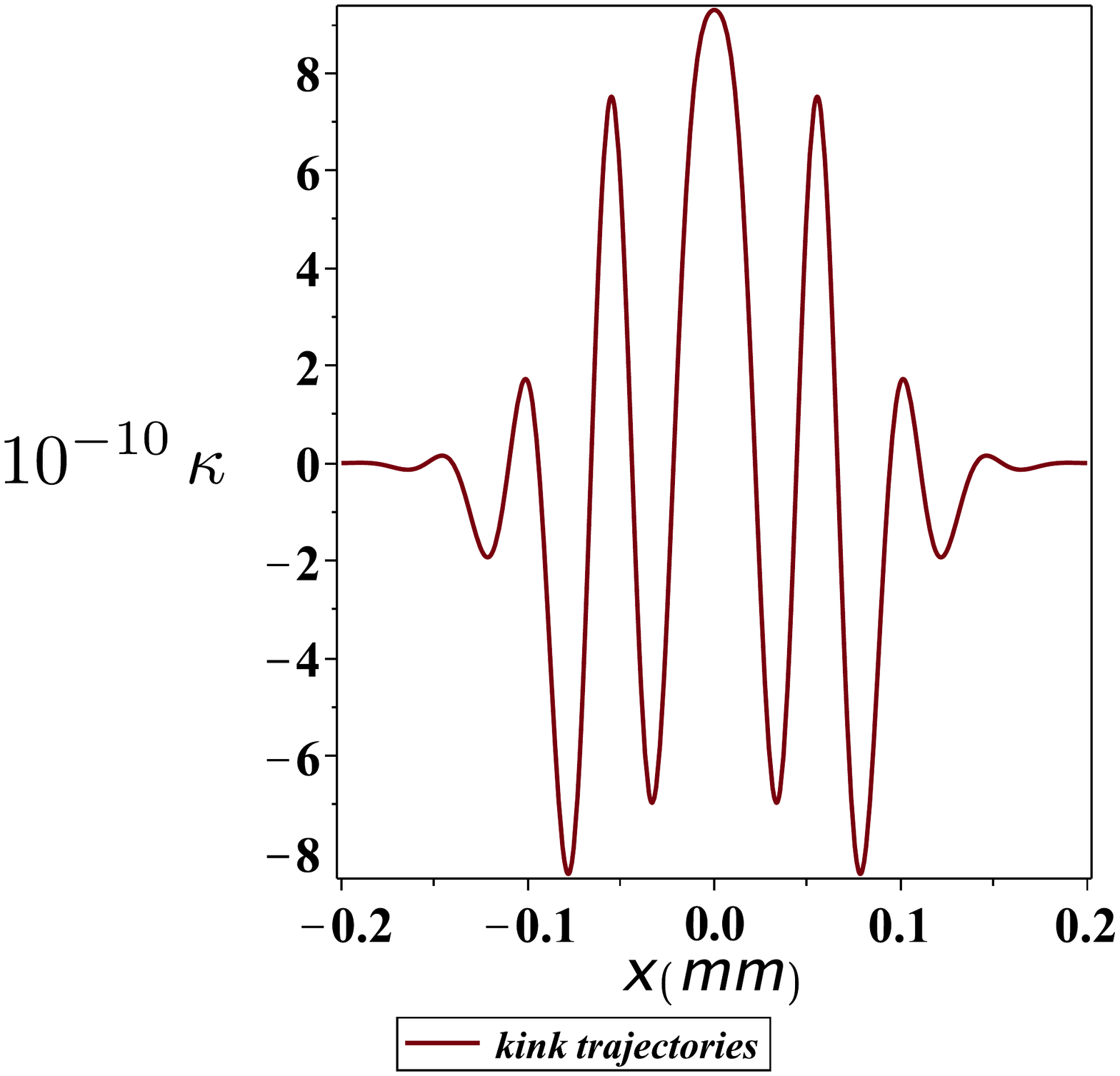}
\includegraphics[width=6 cm]{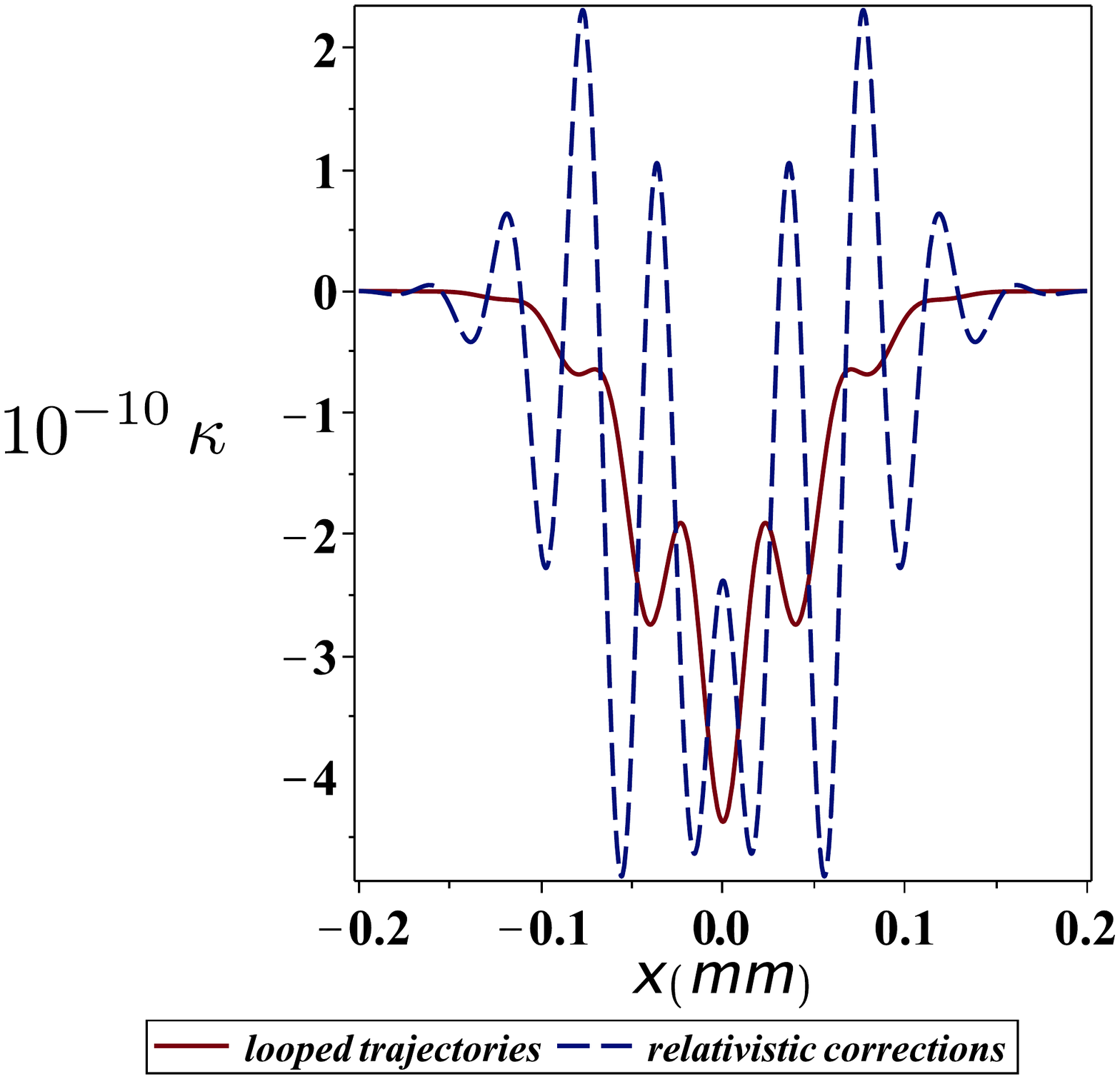}
\caption{The Sorkin parameter due to 3 different contributions for an electron going through a double-slit. The electron has Broglie wavelength $50\,$pm, and the other relevant parameters are $d=272\,$nm, $\beta=62/2\,$nm, $\sigma_0=62\,$nm, and the source-slits (slits-screen) distance is 30.5 cm (24 cm). For these parameters, the inter-slit transit time $\epsilon$ is $0.1$ ns. The variable $x$ is plotted in mm. }
\label{mrankel}
\end{figure}

The Sorkin parameter is evaluated as shown in Eq. (\ref{sorkinDef2}). Plots of the three scenarios are shown in figure \ref{mrank}, in which the parameters referring to a neutron were taken from Ref. \cite{Zeilinger} -- the relativistic corrections were evaluated numerically. Clearly the kink trajectories contribute more significantly, while the contributions from relativistic corrections and looped trajectories are of comparable magnitude.  In figure \ref{mrankel} similar computations were carried out for an electron, in which one can see this hierarchy is such that the kink contributions are still the largest by about 1 order of magnitude in comparison with looped path contributions, which compete with the relativistic corrections.

\section{Conclusions and final remarks}
\label{SectionV}

The study of non-classical or ``exotic'' paths contributions to interferometry of light and matter waves evolved along the endeavours to measure  deviations from the Born's rule and the superposition principle in interference experiments.
Whilst in theory the answer is very simple, as encoded in the Feynman path integral formalism, the boundary conditions involved in the computation of multi-slit diffractions with exotic paths are overwhelmingly difficult and require sophisticated  computational resources. Moreover, for light waves, the absence of a time-dependent formalism to calculate single particle diffraction and issues with photon localization begs for an effective description of the problem.
For example, under the scalar wave approximation the propagation of light is described by the Helmholtz equation subjected to adequate boundary conditions. Thus an effective propagator that satisfies the Fresnel-Huygens principle can be used to compute non-classical trajectories using numerical integration and resource intensive FDTD simulations. Orders of magnitude predicted by theoretical predictions are valuable guides for experimentalists and are worthwhile  exploring. Using an effective double Gaussian approximation for describing type-I SPDC biphotons and the Fresnel approximation to build an effective propagator, we have computed the leading order contributions for biphoton interference in double and triple-slit setups. We have obtained that the Sorkin parameter $\kappa$ can be of order 
$10^{-5}$, which is one order of magnitude larger than typical photon experiments to determine $\kappa$. We have found that 
the spatial correlations encoded in the double Gaussian parameters may play a significant r\^ole in the double-slit setup. Moreover, we have explicitly demonstrated that our simple mathematical model, which can be evaluated using Maple$^\copyright$, reproduces the order of magnitude of the Sorkin parameter for matter waves, such as  the electron, for the same set of parameters used in other simulations. Finally we have addressed the question about the hierarchy of contributing non-classical paths to the Sorkin parameter. We found that kink-shaped paths are of course the leading contributions and that looped paths can contribute with the same order of magnitude as relativistic corrections to the propagator.

\section*{Acknowledgements}
FCVB and CHSV thank Coordenação de Aperfeiçoamento de Pessoal de N{\'i}vel Superior (CAPES). IGP thanks Conselho Nacional de Desenvolvimento Cient{\'i}fico e Tecnol{\'o}gico (CNPq), Grant No. 307942/2019-8. JBA thanks CNPq for the Grant No. 150190/2019-0, and MS thanks CNPq for the Grant No. 303482/2017.

\section*{Appendix A: Biphoton Classical WaveFunction Constants}
\label{apclass}

The general form of a classical trajectory wavefunction at the screen (see Section \ref{SectionIII}) is

\begin{align}
\psi_{i} =A\exp{\left[C_{i}(r,q)+i\alpha_{i}(r,q)\right]}\label{classicalpsi}
\end{align}
with
\begin{align}
C&\equiv c_1 r^2 +c_2 q^2+c_3 r+c_4 q+c_5\nonumber\\
\alpha&\equiv a_1 r^2 +a_2 q^2+a_3 r+a_4 q+a_5,
\end{align}
please note we have intentionally dropped the index $i$, the reason for it will be clear below. The coefficients are given by
\begin{equation}\label{c1}
c_1=-\bigg(\frac{4\pi^2}{\lambda^2 c^2 \tau^2} \bigg) \mathfrak{Re}\bigg[\frac{1}{\omega}\bigg],
\end{equation}
\begin{equation}
c_2=-\bigg(\frac{4\pi^2}{\lambda^2 c^2 \tau^2} \bigg)\mathfrak{Re}\bigg[\frac{1}{\Sigma}\bigg],
\end{equation}
\begin{equation}\label{c3}
c_3= -\frac{\pi }{\lambda c \tau } \bigg( \frac{d_1+d_2}{\beta^2} \bigg)\mathfrak{Im}\bigg[\frac{1}{\omega}\bigg],
\end{equation}
\begin{equation}
c_4=\frac{ \pi }{\lambda c \tau }\bigg( \frac{d_1-d_2}{\beta^2} \bigg) \mathfrak{Im}\bigg[\frac{1}{\Sigma}\bigg],
\end{equation}
\begin{align}\label{c5}
c_5=&-\frac{(d_1^2+d_2^2)}{8 \beta^2}-\frac{(d_1+d_2)^2}{4 \beta^4} \mathfrak{Re} \bigg[\frac{1}{4\omega}\bigg]+\\ &+\frac{(-d_1+d_2)^2}{ \beta^4}\mathfrak{Re}\Bigg[\frac{1}{4 \Sigma}\Bigg],
\end{align}
\begin{equation}
a_1=\frac{2 \pi}{\lambda c t}+\bigg(\frac{4\pi^2}{\lambda^2 c^2 \tau^2} \bigg) \mathfrak{Im}\bigg[\frac{1}{\omega}\bigg],
\end{equation}
\begin{equation}
a_2=\frac{2 \pi}{\lambda c t}+\bigg(\frac{4\pi^2}{\lambda^2 c^2 \tau^2} \bigg) \mathfrak{Im}\bigg[\frac{1}{\Sigma}\bigg],
\end{equation}
\begin{equation}\label{a3}
a_3= -\frac{\pi }{\lambda c \tau }  \frac{(d_1+d_2)}{\beta^2}\mathfrak{Re}\bigg[\frac{1}{\omega}\bigg] ,
\end{equation}
%
\begin{equation}
a_4=\frac{ \pi }{\lambda c \tau }\frac{(d_1-d_2)}{\beta^2}  \mathfrak{Re}\bigg[\frac{1}{\Sigma}\bigg],
\end{equation}
and $a_5$ is given by
\begin{equation}
a_5=\theta+\zeta,
\end{equation}
where
\begin{equation}\label{theta}
\begin{split}
\theta=& - \frac{(d_1+d_2)^2}{4 \beta^4} \mathfrak{Im} \bigg[\frac{1}{4\omega}\bigg]\\ &+ \frac{(-d_1+d_2)^2}{ \beta^4}  \mathfrak{Im}\Bigg[\frac{1}{4 \Sigma}\Bigg],
\end{split}
\end{equation}

\begin{equation}
\zeta =\frac{1}{2}\arctan \bigg(\frac{\mathfrak{Im}[\omega\Sigma]}{\mathfrak{Re}[\omega\Sigma]}\bigg),
\end{equation}
in which
\begin{equation}
\omega = \frac{1}{\beta^{2}}+\frac{2\pi}{I\lambda c \tau}+\frac{1}{\Omega^2+I\frac{ \lambda c t}{ 2\pi \Omega}},
\end{equation}
\begin{equation}
\Sigma = \frac{1}{ \beta^{2}}+\frac{2\pi}{I\lambda c \tau}+\frac{1}{\sigma^2+I \frac{\lambda c t}{ 2\pi \sigma}}.
\end{equation}

The amplitude, which is the same for all classical trajectories, is given by
\begin{equation}\label{A}
A= \bigg(\frac{\pi}{ \lambda c \tau }\bigg) \frac{1}{\sqrt{\pi \vert\omega\vert \vert\Sigma\vert \vert\sigma+ i\alpha/ \sigma\vert \vert\Omega +i\alpha/ \Omega\vert}},
\end{equation}
where $\alpha=\lambda ct/2\pi$.

Now that the coefficients were stated, the index $i$ in Eq. (\ref{classicalpsi}) will tell us which values of $d_{1,2}$ one should take in Eq. (\ref{c3}-\ref{c5}, \ref{a3}-\ref{theta}). As $i=\{uu,dd,ud,du\}$, the table \ref{tableA} displays the choices of $d_{1,2}$  corresponding to each of the four classical contributions. 
\begin{table}[h]
\begin{centering}
\begin{tabular}{|r|r|r|}
\hline 
$i$ & $d_{1}$ & $d_{2}$\tabularnewline
\hline 
\hline 
$uu$ & $-d$ & $-d$\tabularnewline
\hline 
$dd$ & $d$ & $d$\tabularnewline
\hline 
$ud$ & $-d$ & $d$\tabularnewline
\hline 
$du$ & $d$ & $-d$\tabularnewline
\hline 
\end{tabular}
\par\end{centering}
\caption{Choices of $d_{1,2}$ corresponding to the biphoton's four classical
trajectories.}
\label{tableA}
\end{table}

In our numerical evaluations for the plots, since we are interested in the interference pattern, we have disregarded the factor $\sqrt{\frac{\tilde{m}}{2\pi i\hbar(t-t’)}}$ in the propagator of Eq. (16).

\section*{Appendix B: Biphoton Non Classical Wavefunction Constants}
\label{apexotib}
The general form of the non-classical trajectory wavefunction at the screen (See Section \ref{SectionIV} A.) is given by
\bq
\psi_{\text{nc}(a_i,b_i)} (x_1,x_2) &=& A_{\text{nc}(a_i,b_i)}\exp\Big[C_{\text{nc}(a_i,b_i)}(x_1,x_2) \nonumber \\ &+& i\,\, \alpha_{\text{nc}(a_i)}(x_1,x_2)\Big],
\eq%
where the coefficients   $C_{\text{nc}(a_i,b_i)}(x_1,x_2)$ and $\alpha_{\text{nc}(a_i,b_i)}(x_1,x_2)$, omitting the $i$ index, have the general form
\begin{align}
C_{\text{nc}(a,b)}&\equiv \bar{c}_1 x_1^2 +\bar{c}_2 x_2^2+\bar{c}_3 x_1 x_2+\bar{c}_4 x_1+\bar{c}_5 x_2 + \bar{c}_6\nonumber\\
\alpha_{\text{nc}(a,b)}&\equiv \bar{a}_1 x_1^2 +\bar{a}_2 x_2^2+\bar{a}_3 x_1 x_2+\bar{a}_4 x_1+\bar{a}_5 x_2 +\bar{a}_6.\label{alphaCex}
\end{align}%
The coefficients are given by
\begin{equation}
\begin{split}
\bar{c}_1=-\bigg(\frac{\pi^2}{\lambda^2 c^2 \tau^2} \bigg)\mathfrak{Re}\bigg[\frac{1}{\chi_1}\bigg],
\end{split}
\end{equation}%

\begin{equation}
\bar{c}_2=-\bigg(\frac{\pi}{\lambda c \tau} \bigg)^2\mathfrak{Re}\bigg[\frac{1}{\chi_2}\bigg]+\bigg(\frac{\pi^2\Theta}{2\lambda^2 c^2 \tau\epsilon} \bigg)^2\mathfrak{Re}\bigg[\frac{1}{\chi_1\chi_2^2\chi_3^2}\bigg],
\end{equation}%

\begin{equation}
\bar{c}_3=- \frac{\pi^3}{\lambda^3 c^3  \tau^2 \epsilon} \mathfrak{Im}\bigg[\frac{1}{\chi_1\chi_2\chi_3}\bigg],
\end{equation}

\begin{align}\label{c4b}
\bar{c}_4=&\frac{\pi^2 d_3}{2\lambda c  \tau \beta^2} \mathfrak{Im}\bigg[\frac{1}{\chi_1}\bigg]-\frac{\pi^2\Theta}{4\lambda^2 c^2  \tau \beta^2} \bigg(\frac{d_1}{\tau}+\frac{d_2}{\epsilon}\bigg)\mathfrak{Re}\bigg[\frac{1}{\chi_1\chi_2\chi_3}\bigg]\nonumber\\
& +\frac{\pi^2 \Theta^2 d_1}{8\lambda^2 c^2  \tau^2 \beta^2} \mathfrak{Re}\bigg[\frac{1}{\chi_1\chi_2\chi_3^2}\bigg],
\end{align}%
\begin{equation}
\begin{split}
\bar{c}_5=&\frac{\pi d_2}{2\lambda c  \tau \beta^2} \mathfrak{Im}\bigg[\frac{1}{\chi_2}\bigg] -\frac{\pi^3 \Theta^2 }{8\lambda^3 c^3 \tau \epsilon \beta^2} \bigg(\frac{d_1}{\tau}+\frac{d_2}{\epsilon}\bigg)\mathfrak{Im}\bigg[\frac{1}{\chi_1\chi_2^2\chi_3^2}\bigg]\\ & +\frac{\pi^3 \Theta^3 d_1}{16\lambda^3 c^3   \tau \epsilon^2\beta^2} \mathfrak{Im}\bigg[\frac{1}{\chi_1\chi_2^2\chi_3^3}\bigg]-\frac{\pi^2 d_3}{4\lambda^2 c^2  \tau \epsilon \beta^2} \mathfrak{Re}\bigg[\frac{1}{\chi_1\chi_2\chi_3}\bigg],
\end{split}
\end{equation}

\textbf{\begin{equation}\label{c6b}
\begin{split}
\bar{c}_6=&\frac{1}{8\beta^2} (d_1^2+d_2^2+d_3^2)+ \frac{1}{16 \beta^4} \mathfrak{Re}\bigg[\frac{d_1^2}{\chi_3} +\frac{d_2^2}{\chi_2} +\frac{d_3^2}{\chi_1} \bigg]\\ &+ \frac{\pi \Theta d_3}{16 \lambda c \beta^4} \bigg(\frac{d_1}{\tau}+\frac{d_2}{\epsilon}\bigg)\mathfrak{Im}\bigg[\frac{1}{\chi_1\chi_2\chi_3}\bigg]\\ &+ \frac{\pi^2 \Theta^3 d_1}{64 \lambda^2 c^2 \epsilon \beta^4} \bigg(\frac{d_1}{\tau}+\frac{d_2}{\epsilon}\bigg)\mathfrak{Re}\bigg[\frac{1}{\chi_1\chi_2^2\chi_3^3}\bigg]\\ &-\frac{\pi^2 \Theta^2}{64 \lambda^2 c^2 \beta^4}\bigg(\frac{d_1}{\tau}+\frac{d_2}{\epsilon}\bigg)^2\mathfrak{Re}\bigg[\frac{1}{\chi_1\chi_2^2\chi_3^2}\bigg]\\ &  -\frac{\pi d_1 d_3 \Theta^2}{32 \lambda c \epsilon \beta^4}\mathfrak{Im}\bigg[\frac{1}{\chi_1\chi_2\chi_3^2}\bigg] -\frac{\pi^2 d_1^2 \Theta^4}{4\cdot 64 \lambda^2 c^2 \epsilon^2 \beta^4}\mathfrak{Re}\bigg[\frac{1}{\chi_1\chi_2^2\chi_3^4}\bigg],
\end{split}
\end{equation}}%
\begin{equation}
\begin{split}
\bar{a}_1=& \frac{\pi}{\lambda c \tau}+ \bigg(\frac{\pi^2}{\lambda^2 c^2 \tau^2} \bigg)\mathfrak{Im}\bigg[\frac{1}{\chi_1}\bigg],
\end{split}
\end{equation}

\begin{align}
\bar{a}_2=&\frac{\pi}{\lambda c \tau}+ \bigg(\frac{\pi^2}{\lambda^2 c^2 \tau^2} \bigg)\mathfrak{Im}\bigg[\frac{1}{\chi_2}\bigg]\nonumber\\
&-\bigg(\frac{\pi^2\Theta}{2\lambda^2 c^2 \tau\epsilon} \bigg)^2\mathfrak{Im}\bigg[\frac{1}{\chi_1\chi_2^2\chi_3^2}\bigg],
\end{align}

\begin{equation}
\bar{a}_3=- \frac{\pi^3}{\lambda^3 c^3  \tau^2 \epsilon} \mathfrak{Re}\bigg[\frac{1}{\chi_1\chi_2\chi_3}\bigg],
\end{equation}

\begin{equation}\label{a3b}
\begin{split}
\bar{a}_4=&\frac{\pi^2 d_3}{2\lambda c  \tau \beta^2} \mathfrak{Re}\bigg[\frac{1}{\chi_1}\bigg]-\frac{\pi^2\Theta}{4\lambda^2 c^2  \tau \beta^2} \bigg(\frac{d_1}{\tau}+\frac{d_2}{\epsilon}\bigg)\\ & \times\mathfrak{Im}\bigg[\frac{1}{\chi_1\chi_2\chi_3}\bigg] -\frac{\pi^2 \Theta^2 d_1}{8\lambda^2 c^2  \tau^2 \beta^2} \mathfrak{Im}\bigg[\frac{1}{\chi_1\chi_2\chi_3^2}\bigg],
\end{split}
\end{equation}

\begin{equation}
\begin{split}
\bar{a}_5=&\frac{\pi d_2}{2\lambda c  \tau \beta^2} \mathfrak{Re}\bigg[\frac{1}{\chi_2}\bigg] +\frac{\pi^3 \Theta^2 }{8\lambda^3 c^3\tau \epsilon  \beta^2} \bigg(\frac{d_1}{\tau}+\frac{d_2}{\epsilon}\bigg)\\ & \times \mathfrak{Re}\bigg[\frac{1}{\chi_1\chi_2^2\chi_3^2}\bigg] +\frac{\pi^3 \Theta^3 d_1}{16\lambda^3 c^3   \tau \epsilon^2 \beta^2} \mathfrak{Re}\bigg[\frac{1}{\chi_1\chi_2^2\chi_3^3}\bigg]\\ & +\frac{\pi^2 d_3}{4\lambda^2 c^2   \tau \epsilon \beta^2} \mathfrak{Im}\bigg[\frac{1}{\chi_1\chi_2\chi_3}\bigg]
\end{split}
\end{equation}

and $\bar{a}_6=\bar{\theta}+\bar{\zeta}$
\begin{equation}
\begin{split}
\bar\theta=&- \frac{1}{16 \beta^4} \mathfrak{Im}\bigg[\frac{d_1^2}{\chi_3} +\frac{d_2^2}{\chi_2} +\frac{d_3^2}{\chi_1} \bigg]+ \frac{\pi \Theta d_3}{16 \lambda c \beta^4} \bigg(\frac{d_1}{\tau}+\frac{d_2}{\epsilon}\bigg)\\ & \times \mathfrak{Re}\bigg[\frac{1}{\chi_1\chi_2\chi_3}\bigg]- \frac{\pi^2 \Theta^3 d_1}{64 \lambda^2 c^2 \epsilon \beta^4} \bigg(\frac{d_1}{\tau}+\frac{d_2}{\epsilon}\bigg)\mathfrak{Im}\bigg[\frac{1}{\chi_1\chi_2^2\chi_3^3}\bigg]\\ &+\frac{\pi^2 \Theta^2}{64 \lambda^2 c^2 \beta^4}\bigg(\frac{d_1}{\tau}+\frac{d_2}{\epsilon}\bigg)^2\mathfrak{Im}\bigg[\frac{1}{\chi_1\chi_2^2\chi_3^2}\bigg]-\frac{\pi d_1 d_3 \Theta^2}{32 \lambda c \epsilon \beta^4}\\ & \times\mathfrak{Re}\bigg[\frac{1}{\chi_1\chi_2\chi_3^2}\bigg] +\frac{\pi^2 d_1^2 \Theta^4}{4\cdot 64 \lambda^2 c^2 \epsilon^2 \beta^4}\mathfrak{Im}\bigg[\frac{1}{\chi_1\chi_2^2\chi_3^4}\bigg],\label{thetab}
\end{split}
\end{equation}

\begin{equation}
\begin{split}
\bar{\zeta}=&-\frac{1}{2}\arctan \bigg(\frac{\mathfrak{Re}[\chi_1\chi_2\chi_3]}{\mathfrak{Im}[\chi_1\chi_2\chi_3]}\bigg),
\end{split}
\end{equation}
where

\begin{equation}
\chi_1 = \frac{1}{2 \beta^{2}}+\frac{\pi}{I \lambda c \epsilon}+\frac{\pi}{I \lambda c \tau} +\frac{\pi}{\lambda^2 c^2 \epsilon^2 \chi_3}+\frac{\pi^2 \Theta^2}{4\lambda^2 c^2 \epsilon^2\chi_2\chi^2_3},
\end{equation}

\begin{equation}
\begin{split}
\chi_2 =& \frac{1}{ 2\beta^{2}}+\frac{\pi}{I \lambda c \tau}+\frac{1}{\Omega^2+I\frac{ \lambda c t}{ 2\pi }}+\frac{1}{\sigma^2+I\frac{ \lambda c t}{ 2\pi }} -\frac{\Theta^2}{4\chi_3},
\end{split}
\end{equation}

\begin{equation}
\chi_3= \frac{1}{ 2\beta^{2}}+\frac{\pi}{I \lambda c \epsilon}+\frac{1}{\Omega^2+I\frac{ \lambda c t}{ 2\pi }}+\frac{1}{\sigma^2+I\frac{ \lambda c t}{ 2\pi }},
\end{equation}
where $\Theta=\bigg(\frac{1}{\Omega^2+i I\frac{ \lambda c t}{ 2\pi }}- \frac{1}{\sigma^2+I\frac{ \lambda c t}{ 2\pi }}\bigg)$.

The amplitude is given by
\begin{equation}\label{Anc}
\begin{split}
A_{\text{nc}(a,b)}= \frac{1}{\sqrt{\pi \vert\omega\vert \vert\Sigma\vert \vert\sigma+ i\alpha/ \sigma\vert \vert\Omega +i\alpha/ \Omega\vert}}  \frac{\pi^{3/2}}{\lambda c \sqrt{\epsilon \tau \vert \chi_1 \chi_2 \chi_3 \vert}}.
\end{split}
\end{equation}
where $\alpha=\lambda ct/2\pi$.

Now that the coefficients were stated, the table \ref{tableAnc} will tell us which values of $d_{1,2,3}$ one should take in Eq. (\ref{c4b}-\ref{c6b}, \ref{a3b}-\ref{thetab}).
\begin{table}[h]
\begin{centering}
\begin{tabular}{|r|r|r|r|}
\hline 
$\psi_{\text{nc}}$ & $d_{1}$ & $d_{2}$ & $d_{3}$\tabularnewline
\hline 
\hline 
$\psi_{\text{nc}(a_1)}$ & $-d$ & $-d$ & $d$\tabularnewline
\hline 
$\psi_{\text{nc}(a_3)}$ & $d$ & $d$ & $-d$\tabularnewline
\hline 
$\psi_{\text{nc}(b_1)}$ & $-d$ & $d$ & $d$\tabularnewline
\hline 
$\psi_{\text{nc}(b_3}$ & $d$ & $-d$ & $-d$\tabularnewline
\hline 
\end{tabular}
\par\end{centering}
\caption{Choices of $d_{1,2,3}$ corresponding to the biphoton's exotic
trajectories.}
\label{tableAnc}
\end{table}

It is clear that $\psi_{\text{nc}(a_2)}$ and $\psi_{\text{nc}(a_4)}$ ($\psi_{\text{nc}(b_2)}$ and $\psi_{\text{nc}(b_4)}$), are  symmetric under particle exchange to $\psi_{\text{nc}(a_1)}$ and $\psi_{\text{nc}(a_3)}$ ($\psi_{\text{nc}(b_1)}$ and $\psi_{\text{nc}(b_3)}$), respectively, as shown in figure \ref{KinkA} and figure \ref{KinkB}. This means that in order to obtain $\psi_{\text{nc}(a_2)}$ and $\psi_{\text{nc}(a_4)}$ ($\psi_{\text{nc}(b_2)}$ and $\psi_{\text{nc}(b_4)}$), we just have to interchange $x_1$ and $x_2$ in $\psi_{\text{nc}(a_1)}$ and $\psi_{\text{nc}(a_3)}$ ($\psi_{\text{nc}(b_1)}$ and $\psi_{\text{nc}(b_3)}$), respectively.

\section*{Appendix C: Wavefunction constants for a massive particle diffracting through a double-slit}
\label{mass}

\subsection*{I. Classical Path Wavefunction Constants}
Considering a double-slit, which has its slits labeled A and B, one can obtain the wavefunction related to the classical propagation solving the Eq. (\ref{mpsi-classical}), with the slit function given by $F_{A,B}=e^{-\frac{(x'' \mp d/2)^2}{2\beta^2}}$. The   wavefunction that describe the particle leaving the source, going through the upper slit (slit A) and reaching the screen can be written as
\begin{equation}
\begin{split}
\psi_{A} (x)=&\frac{1}{\sqrt{\beta\sqrt{\pi}}}\exp\left[-\frac{(x-D_A/2)^2}{2B^2_A}\right]\\ & \times \exp\left(\frac{i mx^2}{2\hbar R_A}- i \Delta_Ax + i\theta_A+i\mu_A\right),
\end{split}
\end{equation}
where,

\begin{equation}
B_A^2(t,\tau)=\frac{\left(\frac{1}{\beta^2} + \frac{1}{b^2}\right)^2+\frac{m^2}{\hbar^2}\left(\frac{1}{\tau}+\frac{1}{r}\right)^2}{\left(\frac{m}{\hbar\tau}\right)^2\left(\frac{1}{\beta^2} + \frac{1}{b^2}\right)},
\end{equation}

\begin{equation}
R_A(t,\tau)=\tau \frac{\left(\frac{1}{\beta^2} + \frac{1}{b^2}\right)^2+\frac{m^2}{\hbar^2}\left(\frac{1}{\tau}+\frac{1}{r}\right)^2}{\left(\frac{1}{\beta^2}+\frac{1}{b^2}\right)+\left(\frac{t}{\sigma_0 b^2}\right)\left(\frac{1}{\tau} + \frac{1}{r}\right)},
\end{equation}

\begin{equation}\label{D}
D_A(t,\tau)=\frac{\left(1+\frac{\tau}{r}\right)}{\left(1+\frac{\beta^2}{b^2}\right)}d,
\end{equation}

\begin{equation}\label{Delta}
\Delta_A(t,\tau)= \frac{\tau \sigma_0^2 d}{2\tau_0\beta^2 B_A^2},
\end{equation}

\begin{equation}\label{thetaA}
\theta_A(t,\tau)= \frac{md^2\left(\frac{1}{\tau}+\frac{1}{r}\right)}{8\hbar\beta^4\left[\left(\frac{1}{\beta^2} + \frac{1}{b^2}\right)^2+\frac{m^2}{\hbar^2}\left(\frac{1}{\tau}+\frac{1}{r}\right)^2\right]},
\end{equation}

\begin{equation}
\mu_A(t,\tau)=-\frac{1}{2}\arctan\left[\frac{t+\tau\left(1+\frac{\sigma_0^2}{\beta^2}\right)}{\tau_0\left(1-\frac{t\tau\sigma_0^2}{\tau_0^2\beta^2}\right)}\right],
\label{muclas}
\end{equation}

\begin{equation}
b^2(t)=\sigma_0^2\left[1+	\left(\frac{t}{\tau_0}\right)^2\right],
\end{equation}

and

\begin{equation}
r(t)=t\left[1+	\left(\frac{\tau_0}{t}\right)^2\right].
\end{equation}

Here, the parameter  $B_A^2(t, \tau)$ is the beam width for the propagation through one slit, $R_A(t,\tau)$ is the radius of curvature of the wavefronts for the propagation through one slit, $b(t)$ is the beam width for the free propagation and $r(t)$ is the radius of curvature of the wavefronts for the free propagation. $D_{A}(t,\tau)$ is the separation between the wavepackets produced in the double-slit. $\Delta_A(t,\tau) x$ is a phase which varies linearly with the transverse coordinate. $\theta_A(t,\tau)$ and $\mu_A(t,\tau)$ are the time dependent phases and they are relevant only if the slits have different widths. $\mu_A(t,\tau)$ is the Gouy phase for the propagation through one slit. $\tau_0= \frac{m\sigma_0^2}{\hbar}$ is the characteristic time for the ‘aging’ of the initial state. The wavefunction for the propagation going through slit B can be obtained replacing $d$ with $-d$ in Eq. (\ref{D}-\ref{thetaA}).

\subsection*{II. Kink Path Wavefunction Constants}

The wavefunction for a massive particle performing a the kink ($k$) trajectory through a double-slit can be computed by solving the  Eq. (\ref{mpsi-non-classical}), with the slit function given by $F_{A,B}=e^{-\frac{(x'' \mp d/2)^2}{2\beta^2}}$. Considering the propagation that goes from the  slit A (upper slit) to the slit B (lower slit), the corresponding wavefunction can be written as

\begin{equation}
\begin{split}
\psi_{AB}=A_{k} \exp{[C_{k}(x)+i\alpha_{k}(x)]},
\end{split}
\end{equation}
where,

\begin{equation}
\begin{split}
C_k& \equiv c'_{1}x^2+c'_{2}x+c'_{3} \\   \alpha_k & \equiv a'_1 x^2 +a'_2 x +a'_3,
\end{split}
\end{equation}

\begin{equation}
c'_{1}=-\frac{m^2}{4 \hbar^2 \tau^2} \mathfrak{Re} \bigg[\frac{1}{\Gamma_3}\bigg],
\end{equation}

\begin{equation}\label{c2l}
c'_{2}=\frac{m d}{4 \hbar \tau \beta^2} \mathfrak{Im} \bigg[ \frac{1}{\Gamma_3}\bigg]-\frac{m^2 d}{8 \hbar^2 \tau \epsilon \beta^2} \mathfrak{Re} \bigg[\frac{1}{\Gamma_2\Gamma_3}\bigg],
\end{equation}

\begin{equation}
\begin{split}
c'_{3}=&-\frac{d^2}{4\beta^2}+\frac{d^2}{16 \beta^4} \mathfrak{Re}\bigg[\frac{1}{\Gamma_2}\bigg]+\frac{d^2}{16 \beta^4} \mathfrak{Re} \bigg[\frac{1}{\Gamma_3}\bigg]\\ & + \frac{d^2}{16 \hbar \epsilon \beta^4} \mathfrak{Im} \bigg[\frac{1}{\Gamma_2\Gamma_3}\bigg]-\frac{d^2}{64 \hbar^2 \epsilon^2 \beta^4} \mathfrak{Im} \bigg[\frac{1}{\Gamma_2^2 \Gamma_3}\bigg],
\end{split}
\end{equation}

\begin{equation}
a'_1=\frac{m}{2 \hbar \tau}+\frac{m^2}{4 \hbar^2 \tau^2} \mathfrak{Im} \bigg[\frac{1}{\Gamma_3}\bigg],
\end{equation}

\begin{equation}\label{a2l}
a'_2=\frac{m d}{4 \hbar \tau \beta^2} \mathfrak{Re} \bigg[\frac{1}{\Gamma_3}\bigg]+\frac{m^2 d}{8 \hbar^2 \tau \beta^2} \mathfrak{Im} \bigg[\frac{1}{\Gamma_2\Gamma_3}\bigg],
\end{equation}
and $a'_3=\theta'+\mu$, where
\begin{equation}
\begin{split}
\theta'=&-\frac{d^2}{16 \beta^4} \mathfrak{Im}\bigg[\frac{1}{\Gamma_2}\bigg]-\frac{d^2}{16 \beta^4} \mathfrak{Im} \bigg[\frac{1}{\Gamma_3}\bigg]\\ & + \frac{ m d^2}{16 \hbar \epsilon \beta^4} \mathfrak{Re} \bigg[\frac{1}{\Gamma_2\Gamma_3}\bigg]+\frac{m^2d^2}{64 \hbar^2 \epsilon^2 \beta^4} \mathfrak{Im} \bigg[\frac{1}{\Gamma_2^2 \Gamma_3}\bigg],
\end{split}
\end{equation}

\begin{equation}
\mu'= -\frac{1}{2} \arctan\bigg(\frac{\mathfrak{Im}[\Gamma_1\Gamma_2\Gamma_3]}{\mathfrak{Re}[\Gamma_1\Gamma_2\Gamma_3]}\bigg),
\end{equation}
in which

\begin{equation}
\Gamma_1= \frac{1}{2 \sigma_0^2}+\frac{I m}{2\hbar t},
\end{equation}

\begin{equation}
\Gamma_2= \frac{1}{2 \beta^2}+\frac{I m}{2\hbar t}+\frac{I m}{2\hbar \epsilon}+ \frac{m^2}{4 \hbar^2 t^2 \Gamma_1},
\end{equation}

\begin{equation}
\Gamma_3= \frac{1}{2 \beta^2}+\frac{I m}{2\hbar \epsilon}+\frac{I m}{2\hbar \tau}+ \frac{m^2}{4 \hbar^2 \epsilon^2 \Gamma_2}. 
\end{equation}

The amplitude is given by
\begin{equation}
A_{k}= \frac{1}{\sqrt{\sigma_0 \sqrt{\pi}}} \bigg( \frac{m}{2 \hbar} \bigg)^{3/2} \frac{1}{\sqrt{t \epsilon \tau \vert \Gamma_1\Gamma_2\Gamma_3\vert}}.
\end{equation}
The wavefunction $\psi_{BA}$ for the propagation going through slit B, then to slit A, can be obtained replacing $d$ with $-d$ in Eq. (\ref{c2l}) and Eq. (\ref{a2l}).
\subsection*{III. Loop Path Wavefunction Constants}
After some manipulation, the Eq. (\ref{mpsi-loop}) gives the wavefunction that describes a loop propagation ($l$) through a double-slit, with the slit function given by $F_{A,B}=e^{-\frac{(x'' \mp d/2)^2}{2\beta^2}}$. The state for a massive particle propagating first through slit A (upper slit), then loops through slit B (lower), and propagates from slit A to the screen is given by

\begin{equation}
\begin{split}
\psi_{loop-AB}=A_{l} \exp{[C_{l}(x)+i\alpha_{l}(x)]},
\end{split}
\end{equation}
with

\begin{equation}
\begin{split}
C_l & \equiv c''_{1}x^2+c''_{2}x+c''_{3} \\   \alpha_l&\equiv  a''_1 x^2 +a''_2 x +a''_3,
\end{split}
\end{equation}
where

\begin{equation}
c''_{1}=- \frac{m^2}{4 \hbar^2 \tau^2} \mathfrak{Re} \bigg[\frac{1}{\gamma_3}\bigg],
\end{equation}

\begin{equation}\label{c2ll}
\begin{split}
c''_{2}=& -\frac{m d }{4 \beta^2 \hbar \tau} \mathfrak{Im} \bigg[\frac{1}{\gamma_3}\bigg]+\frac{m^2 d }{16 \beta^2 \hbar^2 \tau \epsilon} \mathfrak{Re} \bigg[\frac{1}{\gamma_2 \gamma_3}\bigg]\\ &+\frac{m^3 d }{64 \beta^2 \hbar^3 \tau \epsilon^2} \mathfrak{Im} \bigg[\frac{1}{\gamma_1\gamma_2\gamma_3}\bigg],
\end{split}
\end{equation}

\begin{equation}
\begin{split}
c''_{3}=& -\frac{3d^2}{8 \beta^2}+\frac{d^2 }{16 \beta^4} \mathfrak{Re} \bigg[\frac{1}{\gamma_1}\bigg]+\frac{d^2 }{16 \beta^4} \mathfrak{Re} \bigg[\frac{1}{\gamma_2}\bigg]+\frac{d^2 }{16 \beta^4} \mathfrak{Re} \bigg[\frac{1}{\gamma_3}\bigg]\\ &-\frac{m^2 d^2}{4^4 \beta^4 \hbar^2  \epsilon^2} \mathfrak{Re} \bigg[\frac{1}{\gamma_1\gamma_2}\bigg]+\frac{m^4 d^2 }{4^6 \beta^4 \hbar^4 \epsilon^2} \mathfrak{Re} \bigg[\frac{1}{\gamma_1^2 \gamma_2^2\gamma_3}\bigg]\\ &-\frac{m^2 d^2}{2^7 \beta^4 \hbar^2  \epsilon^2} \mathfrak{Re} \bigg[\frac{1}{\gamma_1 \gamma_2 \gamma_3}\bigg]-\frac{m d^2}{32 \beta^4 \hbar \epsilon} \mathfrak{Re} \bigg[\frac{1}{\gamma_1\gamma_2}+\frac{1}{\gamma_2\gamma_3}\bigg]\\ &-\frac{m^2 d^2}{4^4 \beta^4 \hbar^2  \epsilon^2} \mathfrak{Re} \bigg[\frac{1}{\gamma_2^2\gamma_3}\bigg]-\frac{m^3 d^2}{2^9 \beta^4 \hbar^3  \epsilon^3} \mathfrak{Im} \bigg[\frac{1}{\gamma_1 \gamma_2^2 \gamma_3}\bigg],
\end{split}
\end{equation}

\begin{equation}
a''_{1}= \frac{m}{2 \hbar \tau} + \frac{m^2}{4 \hbar \tau^2}  \mathfrak{Im} \bigg[\frac{1}{\gamma_3}\bigg],
\end{equation}

\begin{equation}\label{a2ll}
\begin{split}
a''_{2}=& -\frac{m d }{4 \beta^2 \hbar \tau} \mathfrak{Re} \bigg[\frac{1}{\gamma_3}\bigg]+\frac{m^2 d }{16 \beta^2 \hbar \tau \epsilon} \mathfrak{Im} \bigg[\frac{1}{\gamma_2 \gamma_3}\bigg] \\ & +\frac{m^3 d }{64 \beta^2 \hbar^3 \tau \epsilon^2} \mathfrak{Re} \bigg[\frac{1}{\gamma_1 \gamma_2 \gamma_3}\bigg]
\end{split}
\end{equation}
and $a''_3=\theta''+\mu''$, where

\begin{equation}
\begin{split}
\theta''_{3}=& -\frac{3d^2}{8 \beta^2}-\frac{d^2 }{16 \beta^4} \mathfrak{Im} \bigg[\frac{1}{\gamma_1}\bigg]-\frac{d^2 }{16 \beta^4} \mathfrak{Im} \bigg[\frac{1}{\gamma_2}\bigg]+\frac{d^2 }{16 \beta^4} \mathfrak{Im} \bigg[\frac{1}{\gamma_3}\bigg]\\ &+\frac{m^2 d^2}{4^4 \beta^4 \hbar^2  \epsilon^2} \mathfrak{Im} \bigg[\frac{1}{\gamma_1^2\gamma_2}\bigg]-\frac{m^4 d^2 }{4^6 \beta^4 \hbar^4 \epsilon^2} \mathfrak{Im} \bigg[\frac{1}{\gamma_1^2\gamma_2^2 \gamma_3}\bigg]\\ &+\frac{m^2 d^2}{2^7 \beta^4 \hbar^2  \epsilon^2} \mathfrak{Im} \bigg[\frac{1}{\gamma_1\gamma_2\gamma_3}\bigg]+\frac{m d^2}{32 \beta^4 \hbar \epsilon} \mathfrak{Re} \bigg[\frac{1}{\gamma_1\gamma_2}+\frac{1}{\gamma_2\gamma_3}\bigg]\\ &+\frac{m^2 d^2}{4^4 \beta^4 \hbar^2  \epsilon^2} \mathfrak{Im} \bigg[\frac{1}{\gamma_2^2\gamma_3}\bigg]-\frac{m^3 d^2}{2^9 \beta^4 \hbar^3  \epsilon^3} \mathfrak{Re} \bigg[\frac{1}{\gamma_1 \gamma_2^2\gamma_3}\bigg],
\end{split}
\end{equation}

\begin{equation}
\mu''=\frac{1}{2} \arctan\bigg(\frac{\mathfrak{Im}[\gamma_0\gamma_1\gamma_2\gamma_3]}{\mathfrak{Re}[\gamma_0\gamma_1\gamma_2\gamma_3]}\bigg),
\end{equation}
where

\begin{equation}
\gamma_0= \frac{1}{2 \sigma_0^2}+\frac{I m}{2\hbar t},
\end{equation}

\begin{equation}
\gamma_1= \frac{1}{2 \beta^2}+\frac{I m}{2\hbar t}+\frac{I m}{2\hbar \epsilon}+ \frac{m^2}{4 \hbar^2 t^2 \gamma_0},
\end{equation}

\begin{equation}
\gamma_2= \frac{1}{2 \beta^2}+\frac{I m}{\hbar \epsilon}+ \frac{m^2}{4 \hbar^2 \epsilon^2 \gamma_1},
\end{equation}

\begin{equation}
\gamma_3= \frac{1}{2 \beta^2}+\frac{I m}{2\hbar \epsilon}+\frac{I m}{2\hbar \tau}+ \frac{m^2}{4 \hbar^2 \epsilon^2 \gamma_2}. 
\end{equation}
The amplitude is given by
\begin{equation}
A_{l}= \frac{1}{\sqrt{\sigma_0 \sqrt{\pi}}}\bigg(\frac{m^2}{4\hbar^2\sqrt{t\epsilon^2 \tau \vert \gamma_0\gamma_1\gamma_2\gamma_3 \vert}}\bigg).
\end{equation}
The wavefunction $\psi_{loop-BA}$ for the propagation going through slit B, then to slit A, can be obtained replacing $d$ with $-d$ in Eq. (\ref{c2ll}) and Eq. (\ref{a2ll}).

\end{document}